\documentclass[fleqn,usenatbib]{mnras}
\usepackage{newtxtext,newtxmath}
\usepackage[T1]{fontenc}
\usepackage{ae,aecompl}
\usepackage{graphicx}	
\usepackage{amsmath}	
\usepackage{amssymb}	

\newcommand       \be           {\begin{equation}}
\newcommand       \ee           {\end{equation}}

\title[Internal Shocks in Classical Novae]{Internal Shocks from Variable Outflows in Classical Novae}

\author[Steinberg \& Metzger]{
Elad Steinberg$^{1}$\thanks{E-mail: es3640@columbia.edu},
Brian D.~Metzger$^{1,2}$
\\
$^{1}$Department of Physics and Columbia Astrophysics Laboratory Columbia University, New York, NY 10027, USA\\
$^{2}$Center for Computational Astrophysics, Flatiron Institute, New York, NY 10010, USA\\
}

\date{Accepted XXX. Received YYY; in original form ZZZ}

\pubyear{2015}

\begin{document}
\label{firstpage}
\pagerange{\pageref{firstpage}--\pageref{lastpage}}
\maketitle

\begin{abstract}
We present one-dimensional hydrodynamical simulations including radiative losses, of internal shocks in the outflows from classical novae, to explore the role of shocks in powering multi-wavelength emission from radio to gamma-ray wavelengths.  Observations support a picture in which the initial phases of some novae generate a slow, equatorially-focused outflow (directly from the outer Lagrange point, or from a circumbinary disk), which then transitions to, or is overtaken by, a faster more isotropic outflow from the white dwarf which collides and shocks the slower flow, powering gamma-ray and optical emission through  reprocessing by the ejecta.  However, the common occurence of multiple peaks in nova light curves suggests that the outflow's acceleration need not be monotonic, but instead can involve successive transitions between ``fast" and ``slow" modes.  Such a time-fluctuating outflow velocity naturally can reproduce several observed properties of nova, such as correlated gamma-ray and optical flares, expansion of the photosphere coincident with (though lagging slightly) the peak flare luminosity, and complex time-evolution of spectral lines (including accelerating, decelerating, and merging velocity components).  While the shocks are still deeply embedded during the gamma-ray emission, the onset of $\sim$ keV X-ray and $\sim 10$ GHz radio synchrotron emission is typically delayed until the forward shock of the outermost monolithic shell (created by merger of multiple internal shock-generated shells) reaches a sufficiently low column through the dense external medium generated by the earliest phase of the outburst.
\end{abstract}




\section{Introduction}

A classical nova occurs when a shell of hydrogen-rich material on the surface of a white dwarf, which is accreting from a main sequence binary companion, undergoes runaway nuclear burning \citep{Gallagher&Starrfield78,Prialnik86,Bode&Evans08,Yaron+05,Starrfield+16}.  The resulting energy release leads to the ejection of a shell of mass $\sim 10^{-3}-10^{-7}M_{\odot}$ at high velocities $\sim 500-5000$ km s$^{-1}$ over days to months or longer.  The ejecta contains the original accreted material and its burning products, as well as heavier nuclei dredged up from the underlying white dwarf (e.g. by Kelvin-Helmholtz instabilities at the interface between the inert core and the burning envelope; e.g.~\citealt{Casanova+18}).    

Traditionally, models of novae predicted that their light curves are powered exclusively by radiation which diffuses outwards directly from the hot white dwarf envelope through the expanding ejecta \citep{Gallagher&Starrfield78,Yaron+05}.  Early in the outburst, the ejecta are optically thick, and this luminosity emerges primarily at visual wavelengths.  As the ejecta becomes increasingly transparent with time, the spectral peak shifts into the UV and, ultimately, the soft X-rays, once the white dwarf surface becomes visible (e.g.~\citealt{Schwarz+11,Henze+14}), powered by stable burning of the residual fuel (e.g.~\citealt{Wolf+13}).

However, a number of complications have long afflicted this simple picture.  The optical light curves of many nova show multiple peaks (e.g.~\citealt{Strope+10}) and complex spectral components, suggestive of a number of discrete mass ejection events (e.g.~\citealt{Cassatella+04,Csak+05,Hillman+14}).  The photosphere appears to temporarily expand during flares (e.g.~\citealt{Tanaka+11a,Aydi+19}), behavior often accompanied by the appearance of new absorption line systems (e.g.~\citealt{Jack+17}).  Furthermore, X-ray observations reveal the presence of hot $\gtrsim 10^{7}$ K gas indicative of shocks \citep{OBrien+94,Mukai&Ishida01,Nelson+19}.  However, the prevalance, and energetic importance, of these shocks were not fully appreciated until the discovery by {\it Fermi}/LAT of $\gtrsim 100$ MeV gamma-rays from Galactic classical novae (\citealt{Ackermann+14,Cheung+16,Franckowiak+18}).  

The gamma-rays observed from classical novae are very likely produced by relativistic particles accelerated at shocks internal to the ejecta (e.g.~\citealt{Martin&Dubus13,Metzger+15,Martin+18}).  The high columns of gas ahead and behind the shocks make nova outflows excellent ``calorimeters'' for converting cosmic ray energy into gamma-rays \citep{Metzger+15}.
For typical acceleration efficiencies of at most tens of percent of the shock power going into the energy of relativistic particles (e.g.~\citealt{Caprioli&Spitkovsky14}), the high observed gamma-ray luminosities, $L_{\gamma} \sim 10^{35}-10^{36}$ erg s$^{-1}$, require shocks with kinetic powers approaching the bolometric output of the nova, typically at or near the Eddington luminosity of the white dwarf, $L_{\rm Edd} \sim 10^{38}$ erg s$^{-1}$.  

The high densities in nova outflows also imply that the hot shocked gas will cool rapidly compared to the expansion time, i.e.~the shocks are radiative rather than adiabatic.  The layer of gas behind a radiative shock in this temperature range are often subject to thermal instabilities (\citealt{Chevalier&Imamura82}) as well as dynamical instabilities in the thin shell generated by the gas' compression (\citealt{Vishniac94}).  In conflict with naive expectations based on the 1D jump conditions (which would predict mainly hard $\gtrsim$ keV X-ray emission for $\sim 10^{3}$ km s$^{-1}$ shocks), recent multi-dimensional hydrodynamical simulations of radiative shocks show that these instabilities can reduce the temperature of the bulk of the luminosity to soft X-ray or UV wavelengths \citep{Kee+14,Steinberg&Metzger18}.  However, as we now discuss, such direct emission from shock is anyways not typically visible to an external observer until late times due to absorption by the overlying ejecta.

Early in the nova outburst, when the mass loss rate from the white dwarf is greatest, any UV and X-ray emission from the shocks will be absorbed by gas ahead of the shocks, reprocessing its energy into optical line or continuum emission.  In fact, the required shock powers are so high that it led \citet{Metzger+14} to conclude that a significant fraction of the total optical luminosity of the nova may be powered (indirectly) by shocks (see also \citealt{Metzger+15,Martin+18}) rather than the white dwarf.  This possibility has received strong support from the observed temporal correlations between the optical and gamma-ray light curves (\citealt{Li+17}; Aydi et al., in prep).  High gas densities behind the shocks also imply that primary and secondary relativistic leptons (electrons or positrons) accelerated at the shock transfer their energy through Coulomb scattering to the thermal plasma faster than it can be radiated (e.g.~via Inverse Compton scattering).  This suppresses the non-thermal X-ray emission relative to what one would expect based on a naive downward extrapolation of the $\sim$ GeV photon spectrum measured by {\it Fermi} \citep{Vurm&Metzger18}, consistent with deep upper limits from {\it NuSTAR} on a component of non-thermal X-rays contemporaneous with {\it Fermi} detections \citep{Nelson+19}.      

Further evidence regarding the nature of the shocks in novae comes from late-time ($\gtrsim$ several month to year timescale) radio observations, both of thermal emission from the expanding photo-ionized ejecta (e.g.~\citealt{Ribereiro+14,Cunningham+15}) as well non-thermal synchrotron emission from shocks (e.g.~\citealt{Krauss+11,Weston+16}).  Resolved radio imaging supports a picture in which the initial stages of the nova eruption generate a slow outflow that is geometrically-concentrated in the binary equator, which then transitions to$-$or is overtaken by$-$a faster outflow or wind with more isotropic angular distribution \citep{Sokoloski+08,Chomiuk+14,Linford+15}.  This outflow sequence is consistent with the bipolar ring- and jet-like structures revealed by optical imaging and spectroscopy of nova shells (e.g.~\citealt{Gill&OBrien00,Harman&OBrien03,Woudt+09,Shore13,Ribeiro+13,Shore+13}).
The ``slow" equatorial outflow plausibly arises from an outflow or circumbinary disk which is fed by Roche lobe overflow at velocities $\lesssim$ few 100 km s$^{-1}$ comparable to the binary orbital speed.  In contrast, the ``fast" ejecta likely originates directly from the white dwarf surface, e.g. through a momentum-driven wind (\citealt{Kato&Hachisu94}). 

\begin{figure}
    \centering
    \includegraphics[width=0.5\textwidth]{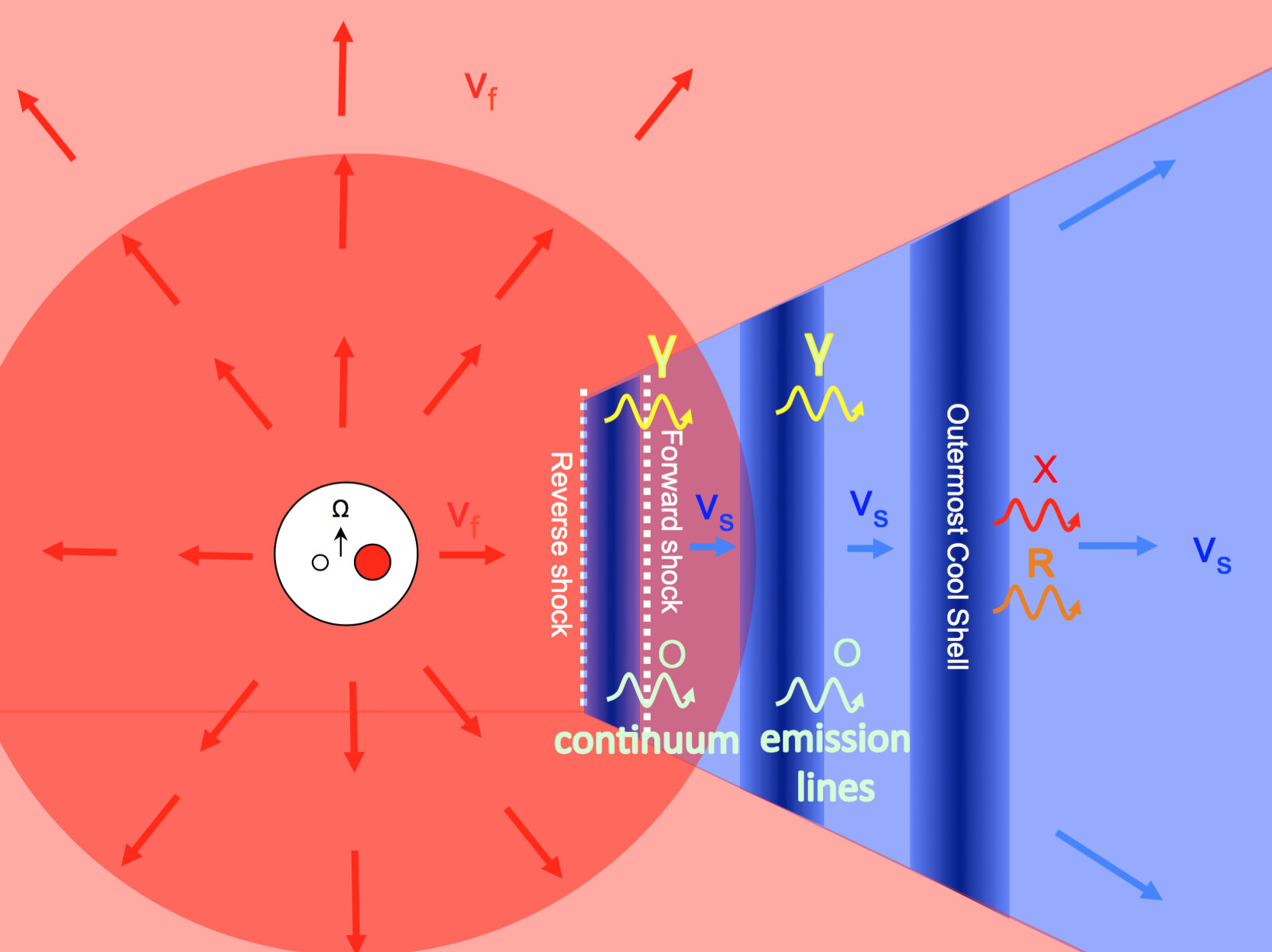}
    \caption{Schematic illustration of time-variable outflows and internal shocks in classical novae, as viewed through the binary equatorial plane.  Our 1D hydrodynamical calculations model the interaction between a time-variable fast outflow from the white dwarf (velocity $v_{\rm f}$), with a slow outflow (velocity $v_{\rm s}$) concentrated in the binary plane.  Each time the outflow mode switches from the slow to fast mode, the resulting collision generates internal shocks which sweep up gas into a cool thin shell.  These radiative shocks generate a correlated gamma-ray and optical flare via ejecta reprocessing of accelerated relativistic particles and thermal UV/X-ray emission, respectively.  Shocks contribute to the continuum optical flux when the shocks are below the photosphere, or to emission lines once the shocks emerge above photosphere.  The dense shells may merge on large scales into a single shell, which continues to propagate into the initial slow outflow which was generated at (or even existing around the binary prior to) the thermonuclear runaway.  X-ray and radio emission from the forward shock become visible once the gas column ahead of it becomes low enough, as typically occurs on a timescale of weeks to months. }
    \label{fig:cartoon}
\end{figure}

Collisions between ``fast'' and ``slow'' outflow components thus provide a natural site for shocks and bipolar collimation in novae.  However, the physical mechanism giving rise to the transitions between these markedly distinct outflow modes remains unclear.  Most previous theoretical work has considered the evolution of the outflow velocity from slow to fast to be monotonic in time (e.g.~\citealt{Metzger+14,Martin+18}).  However, the presence of multiple flares in nova optical and gamma-ray light curves, if interpreted as distinct collision events, requires multiple transitions between fast and slow modes in a single event  (e.g.~ \citealt{Kato&Hachisu09}).  Rather than a single shock, such a scenario would generate an entire `train' of collisions.  At late times, both single- and multiple-transition scenarios will generate a broadly similar global hourglass geometry of a fast bipolar wind confined by the ``waist" of the slow equatorial shell of shock-compressed gas (Fig.~\ref{fig:cartoon}).  However, at early times (e.g.~during the observed gamma-ray flares), these scenarios make distinct predictions for the physical locations of the shocks and their impact on the observed emission.


This work explores the role of internal shocks in nova outflows and their emission using global one-dimensional hydrodynamical simulations which include radiative losses.  1D calculations cannot account for latitudinal or azimuthal variations in the shocks, nor can they resolve important multi-dimensional features of the shock front.  The latter can be approximately accounted for in a sub-grid sense using results gained from previous local multi-dimensional simulations of radiative shocks \citep{Steinberg&Metzger18}.  In this paper we simply point out where multi-dimensional results are likely to have the greatest affects on our conclusions and thus where some caution in the interpretation is warranted.  

This paper is organized as followed.  In \S\ref{sec:numerical} we describe details of our numerical model (see also Appendix \ref{app:numerical}).  In \S\ref{sec:results} we present results of the simulations.  We compare the single- and multiple-transition scenarios, focusing on their abilities to reproduce the observed variability in nova optical light curves as well as the evolution of the photospheric radius and line velocity evolution.  We also present $\gamma-$ray, X-ray, and radio light curves from our models, and briefly discuss implications for dust formation.  In \S\ref{sec:conclusions} we summarize our conclusions.

\section{Description of Numerical Model}
\label{sec:numerical}

We perform one-dimensional hydrodynamical simulations of nova outflows using the moving-mesh hydrodynamical code RICH on a spherical grid \citep{Yalinewich+15}.  Our aim is to capture the shock interaction between fast and slow outflow components, which we expect to occur primarily in the binary equatorial plane (Fig.~\ref{fig:cartoon}).  We are therefore not accounting for additional shocks or emission which could potentially take place in the polar regions due to velocity fluctuations in the fast outflow itself.  The latter could readily be studied in an approximate manner with a similar 1D calculation to that presented here in future work.

As discussed in greater detail below, we specify the time-dependent velocity and mass-loss rate from the central spherical source (white dwarf atmosphere or circumbinary disk) via an inner boundary condition on the radial velocity and density at the radius $r = 5\times 10^{11}$ cm.  We employ a radial grid extending to an outer radius of $2.5\times 10^{15}$ cm with a high enough resolution to capture the large compression behind the shocks.  We assume ideal gas pressure (adiabatic index $\gamma = 5/3$) and include optically-thin radiative losses, using a cooling function calculated assuming collisional ionization equilibrium (CIE) as tabulated in version C17.01 of CLOUDY \citep{Ferland+98,Ferland+17}.  In our fiducial model we adopt the GASS10 solar abundances, but we also explore models using CLOUDY's default {\it nova} abundances to account for the effects of possible enrichment of the ejecta from nuclear burning (\citealt{Starrfield+72}).  

It is reasonable to treat the cooling as being effectively optically-thin for purposes of capturing the dynamical evolution.  This is because the timescale for radial diffusion of the reprocessed optical radiation is much shorter than the dynamical or expansion timescale of the outflow, both on global scales and on smaller scales across the post-shock cooling layer (e.g.~\citealt{Metzger+14}).  Since our simulations do not consider heating of the gas by the outwards diffusion of thermal radiation from the white dwarf or shocks, we place a minimum temperature floor of $T_{\rm min} = 10^{4}$ K similar to that maintained by the radiation through the layers of greatest interest.  Our results are not sensitive to the precise value of the floor, as its main effect is limit the maximum compression ratio of the shocked gas.  Additional details of our numerical set-up are described in Appendix \ref{app:numerical}.  

Mass loss is a key ingredient in novae, which plays an important role in setting the timescale of the eruption by removing mass from the white dwarf envelope \citep{Livio+90,Kato&Hachisu94}.  However, the exact mechanisms giving rise to the unbound outflows, as needed to produce variable internal shocks, are uncertain and not addressed in this work.  Relatively slow outflows, of up to several hundreds of km s$^{-1}$, may be launched from near the binary $L_{2}$ Lagrange point.  These could be driven for instance by frictional drag (e.g.~\citealt{Livio+90,Shankar+91,Lloyd+97}; however, see \citealt{Kato&Hachisu91}), non-axisymmetric gravitational torques from the binary \citep{Pejcha+16a}, or in the outflow of a circumbinary disk formed from matter ejected from $L_{2}$ that initially remains gravitationally bound \citep{Pejcha+16b}.  These slow outflows, with velocities of order the binary escape speed, would plausibly be concentrated in the binary equatorial plane.

Outflows can also occur from the white dwarf directly, with little or no influence from the binary.  These could be driven by radiation pressure acting on the iron opacity bump (e.g.~\citealt{Kato&Hachisu94}), or by any large (near- or super-Eddington) localized energy deposition in the white dwarf envelope, e.g. from radioactive heating or damping of convectively-excited waves (e.g.~\citealt{Quataert+16}).  Outflows launched from deeper locations in the potential well of the white dwarf would be expected to possess higher velocities $\sim 2000-5000$ km s$^{-1}$ and a more spherical geometry than the slower outflows from $L_{2}$.  

Lacking a first principles model, we are motivated on phenomenological grounds to consider two outflow modes characterized by distinct mass-loss rates, $\dot{M}$, and asymptotic velocities, $v_{\rm w}$.  Somewhat arbitrarily (as the precise values will vary between novae), we assume a ``fast'' mode outflow with
\be
\dot{M} = \dot{M}_{\rm f} =  10^{-5}M_\odot\,{\rm week^{-1}}, v_{\rm w} = v_{\rm f} = 1400\,{\rm km\,s^{-1}} 
\ee
and a ``slow'' outflow mode with
\be
\dot{M} = \dot{M}_{\rm s} =  5\cdot10^{-6}M_\odot {\rm week^{-1}}, v_{\rm w} = v_{\rm s} = 200\,{\rm \,km\,s^{-1}}.
\ee
These are spherically-equivalent mass-loss rates.  To obtain the true value, $\dot{M}$ must be multipled by the fraction of the total solid angle covered by each outflow component, $f_{\Omega}$.  For instance, $f_{\Omega} \approx 1$ for the fast spherical outflow, while $f_{\Omega} \approx 0.3$ for the slow equatorially-focused outflow.\footnote{This is consistent with $\sim 30\%$ of nova showing evidence for dust creation along the line of site assuming the dust forms most effectively in the shock-compressed slow equatorial ejecta (\citealt{Derdzinski+17}; see $\S\ref{sec:dust}$).}

\begin{figure}
    \centering
    \includegraphics[width=0.5\textwidth]{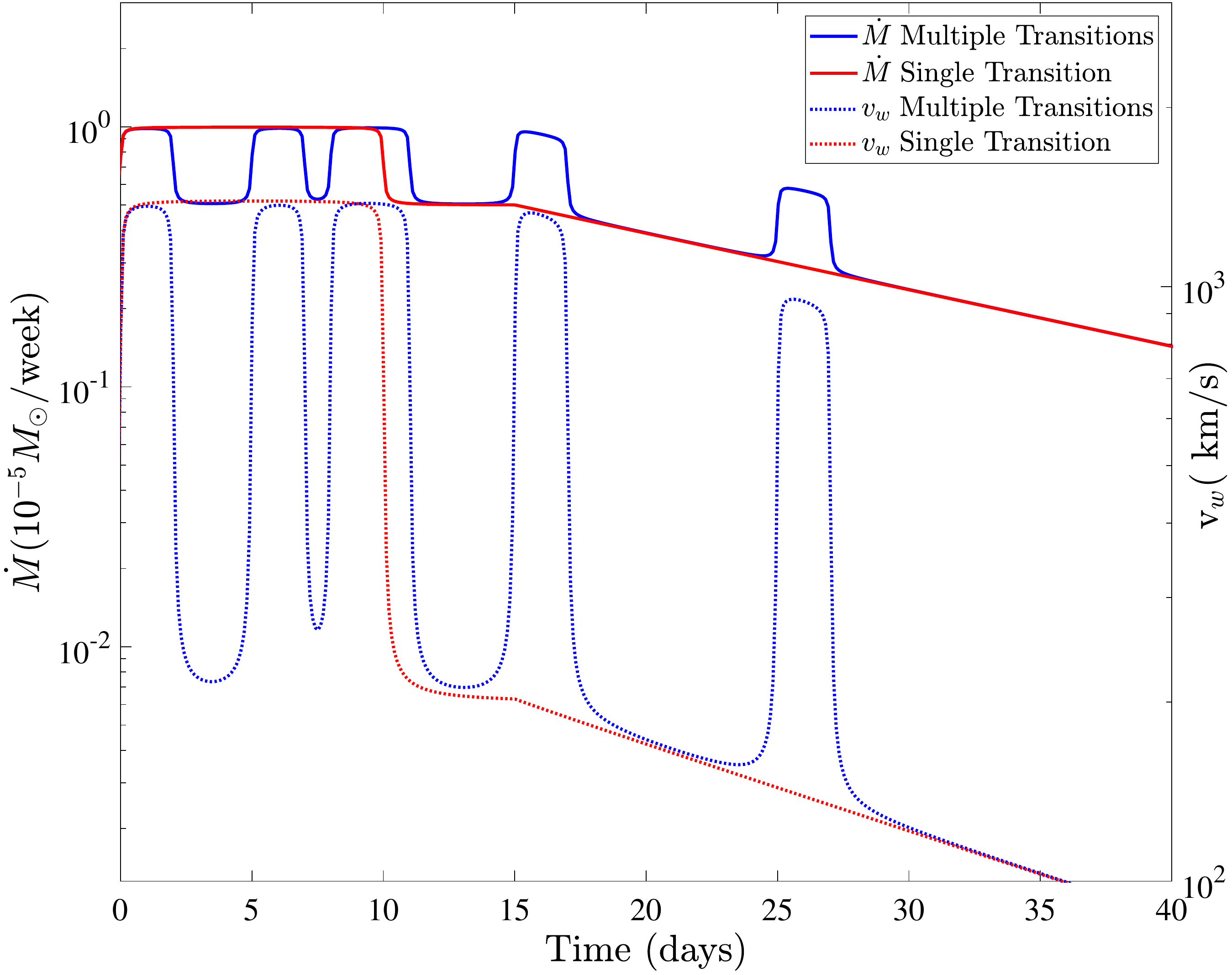}
     \includegraphics[width=0.5\textwidth]{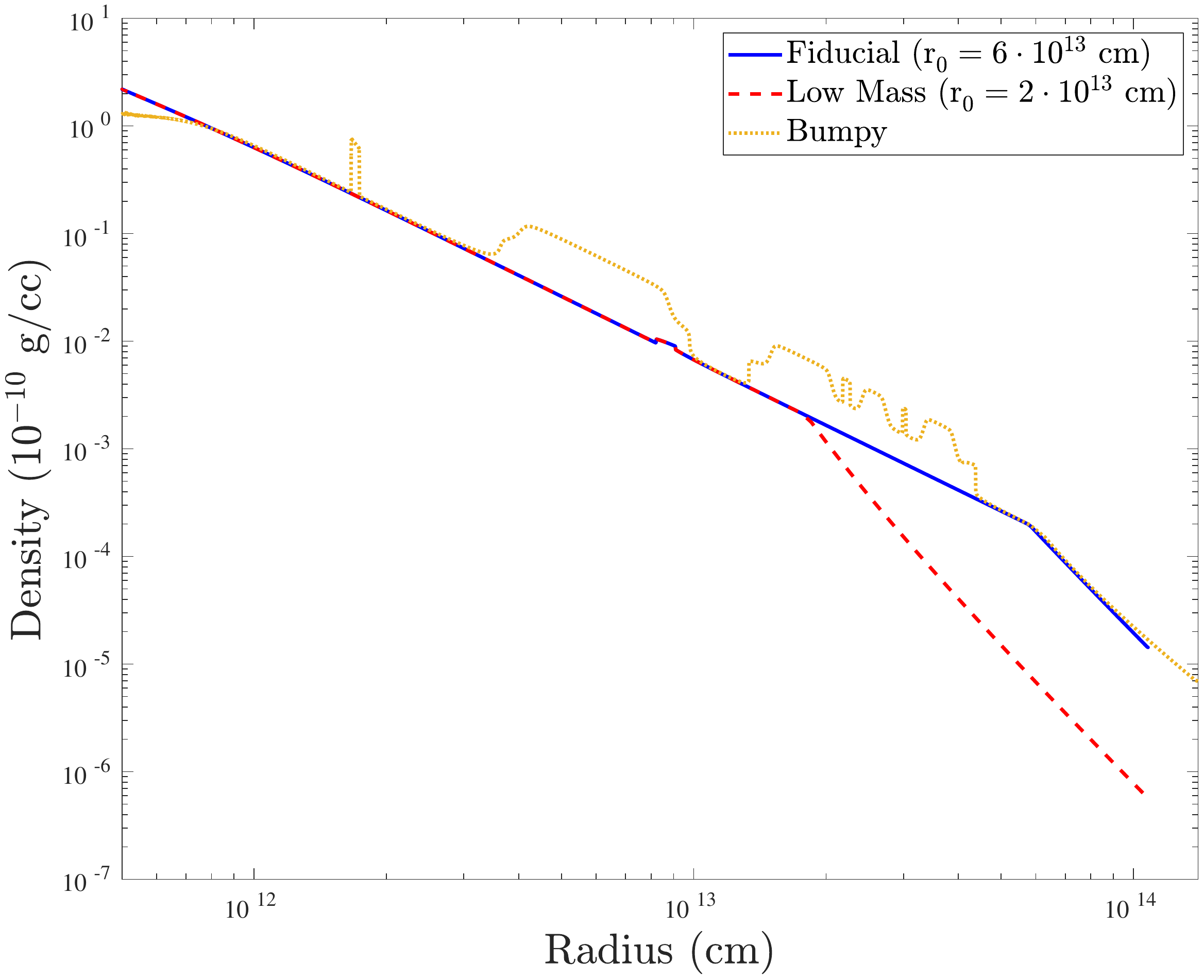}
    \caption{{\it Top:} Time evolution of the mass-loss rate $\dot{M}$ and velocity $v_{\rm w}$ of the nova outflow in our two fiducial models.  The beginning of the simulation ($t = 0$) coincides with the first transition to a fast outflow.  In the ``single transition'' model (solid lines), the nova outflow increases from a slow ($v_{\rm w} = v_{\rm s}$) to a fast ($v_{\rm w} = v_{\rm f}$) outflow only once.  In the ``multiple transition'' model the outflow transitions to the fast component over brief intervals several times over the first week to month.    In both cases the outflow reverts back to the slow mode at late times as we taper $\dot{M}$ and $v_w$ to mimick the fading of the nova outburst. {\it Bottom:}  Here we show the radial density profile $\rho(r)$ surrounding the white dwarf at the beginning of the simulation (first onset of the fast wind; $t = 0$).  At small radii $r < r_0$ the profile $\rho \propto 1/r^{2}$ at $r < r_0$ is that of the initial slow outflow, while at large radii $r > r_0$ the profile steepens $\rho \propto 1/r^{4}$ at $r > r_0$, corresponding to the pre-existing external medium at the onset of the slow outflow.  We consider separate models for two different values of $r_0 = 6\times 10^{13}$ cm (``Fiducial") or $r_0 = 2\times 10^{13}$ cm (``Low Mass"), corresponding to different durations of the initial slow outflow.  We also consider a model (``Bumpy") in which the medium is radially inhomogeneous. }
    \label{fig:inject}
\end{figure}

The top panel of Figure \ref{fig:inject} shows the chosen time evolution of $\dot{M}$ and $v_{\rm w}$ for our two fiducial models (``single transition'' and ``multiple transition'').  In the single transition model (red lines in Fig.~\ref{fig:inject}) the outflow velocity is increased from slow ($v_{\rm w} = v_{\rm s}$) to fast ($v_{\rm w} = v_{\rm f}$) only once at $t \approx 0$, while in the multiple transition model (blue lines in Fig.~\ref{fig:inject}) the outflow velocity fluctuates between slow and fast modes several times.  Local maxima (``flares'') are observed in classical nova light curves on timescales ranging from hours to weeks (e.g.~\citealt{Strope+10,Walter+12,Henze+18,Aydi+19}).  To generate collisions on approximately this timescale, we impose transitions between the slow to the fast mode on a characteristic time interval $\delta t \approx$ hours$-$days and with delays between the transitions of a day to a week.  In our multiple transition model, we space the transitions across a range of $\delta t$, in order to explore its effect on the behavior of the photosphere during flares.  In both models, the outflow velocity $v_w$ returns back to the slow mode at late times, which we further taper along with $\dot{M}$ to mimic the end of the outburst. 

We assume that by the start of our simulations ($t = 0$) the initial slow outflow has already reached a characteristic radius $r_0$, outside of which we truncate the density as $\rho \propto r^{-4}$ (Fig.~\ref{fig:inject}, bottom panel).  Early spectra of novae showing narrow absorption line features indicative of the presence of slowly-expanding material already present around the binary by the time of the nova outburst (e.g.~\citealt{Williams+08}).  The origin of this material is unclear: it may represent the earliest stages of mass ejection after the thermonuclear runaway, or, more speculatively, pre-existing material initially orbiting the binary.  In our ``Fiducial" model we take $r_0 = 6\times 10^{13}$ cm for the radial extent of the pre-existing medium,  in which case the total mass of the ejecta in the fast and slow components (assuming $f_{\Omega} = 1$ for both) are $\approx 1.5\times 10^{-5}M_{\odot}$ and $\approx 5.5\times 10^{-5}M_{\odot}$, consistent with the measured range in classical novae (e.g.~\citealt{Roy+12}).  We also consider a ``Low Mass'' model with $r_0 = 2\times 10^{13}$ cm and thus a lower mass in the slow component by a factor of a few compared to the Fiducial case.  As we will show later, the X-ray and radio light curves of the shocks are sensitive to the properties of the initial external medium.  Finally, we consider a model (``Bumpy") in which the initial slow outflow extends to $r_0 = 6\times 10^{13}$ cm but is radially inhomogeneous.  

Nova outflows are intrinsically multi-dimensional, both on large spatial scales (e.g.~equatorial-polar asymmetry) and on small scales (e.g. the cooling length of the post-shock gas) due to instabilities of the radiative shocks.  Using zoomed-in high-resolution 2D and 3D simulations of head-on colliding dense flows, \citet{Steinberg&Metzger18} demonstrate that the luminosity-weighted temperature of the emission from the gas behind high Mach number radiative shocks is substantially lower than would be guessed from the adiabatic Rankine-Hugoniot jump conditions.  This is due to a combination of obliquities generated at the shock front by the non-linear thin-shell instability \citep{Vishniac94} and the transfer of thermal energy from the hot immediate post-shock gas to cool dense clumps downstream (which radiates energy away more efficiently).  The present 1D treatment of nova outflows will therefore underestimate the thickness of the cool shells of shock-compressed gas and lead to overestimates of the emission temperature (see $\S\ref{sec:Xrays}$).  The former deficiency will not alter our qualitative conclusions, but can lead to artificially high luminosity when two such thin-shells collide ($\S\ref{sec:luminosity}$).

\section{Internal Shocks in Novae}
\label{sec:results}

In this section we present the results of our simulations and our predictions for the optical, gamma-ray, X-ray, and radio emission in comparison to observed nova properties.


\subsection{Single Versus Multiple Transition Scenarios}
\label{sec:luminosity}

We begin with a general discussion of the dynamics of the outflow self-interaction and the resulting optical light curves.  In the single transition scenario, the nova outflow undergoes a single change from a slow to fast outflow at $t \approx 0$ (Fig.~\ref{fig:inject}).  The fast flow quickly catches up to the slow flow, generating a single forward-reverse shock structure that propagates outwards through the slow outflow.  The shocks sweep up mass from both flows into a shell that decelerates in a momentum-conserving fashion \citep{Metzger+14}.  The shell is extremely thin due to radiative losses; following cooling, the gas density increases by a factor of $\propto \mathcal{M}^{2} \sim 10^{4}$ compared to its original unshocked value, where $\mathcal{M} = v_{\rm sh}/c_{\rm s}$ is the Mach number of the shocks, $v_{\rm sh}$ is the shock speed, and $c_{\rm s} \sim 10$ km s$^{-1}$ is the sound speed of the upstream.  

\begin{figure}
    \centering
    \includegraphics[width=0.5\textwidth]{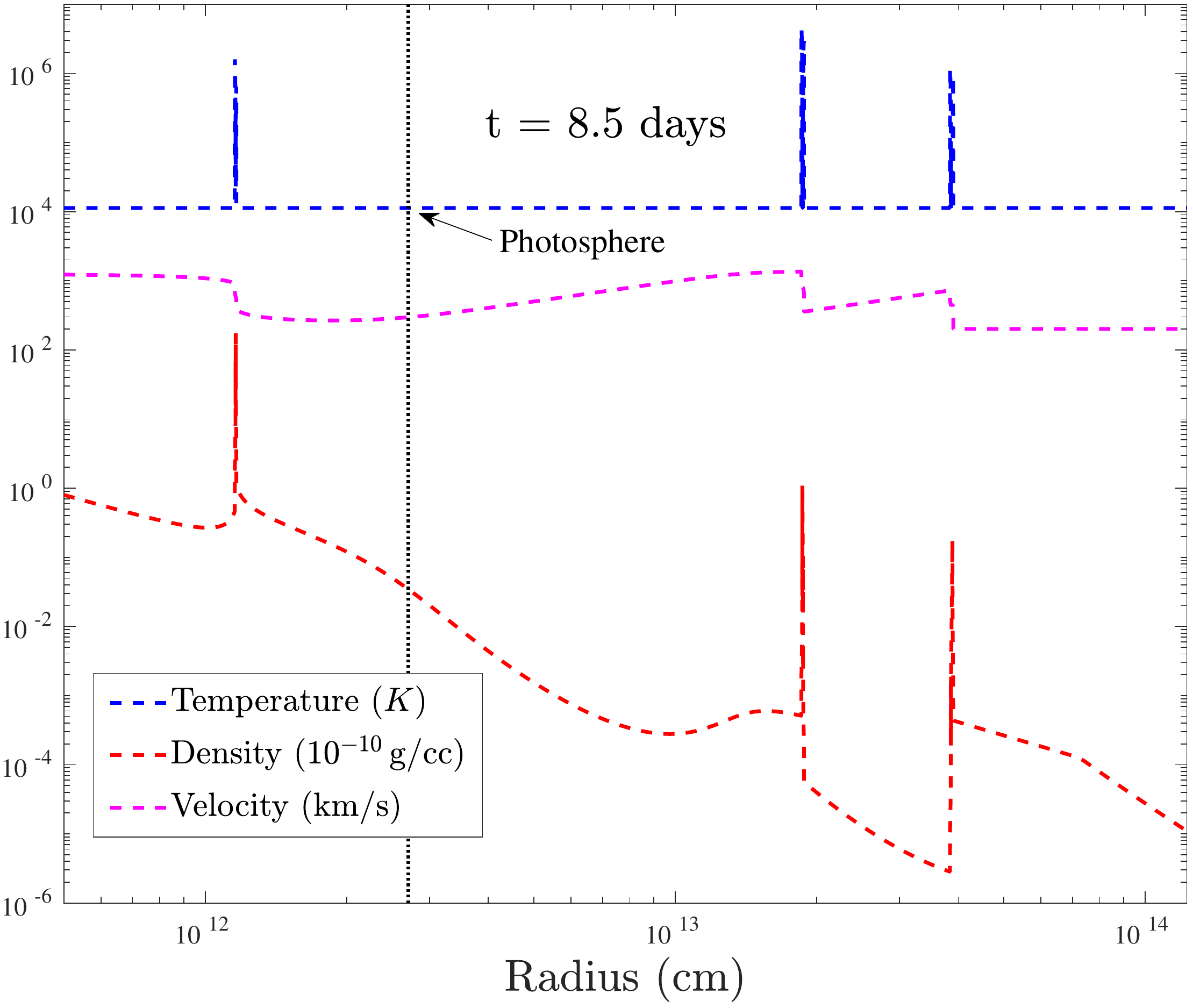}
    \includegraphics[width=0.5\textwidth]{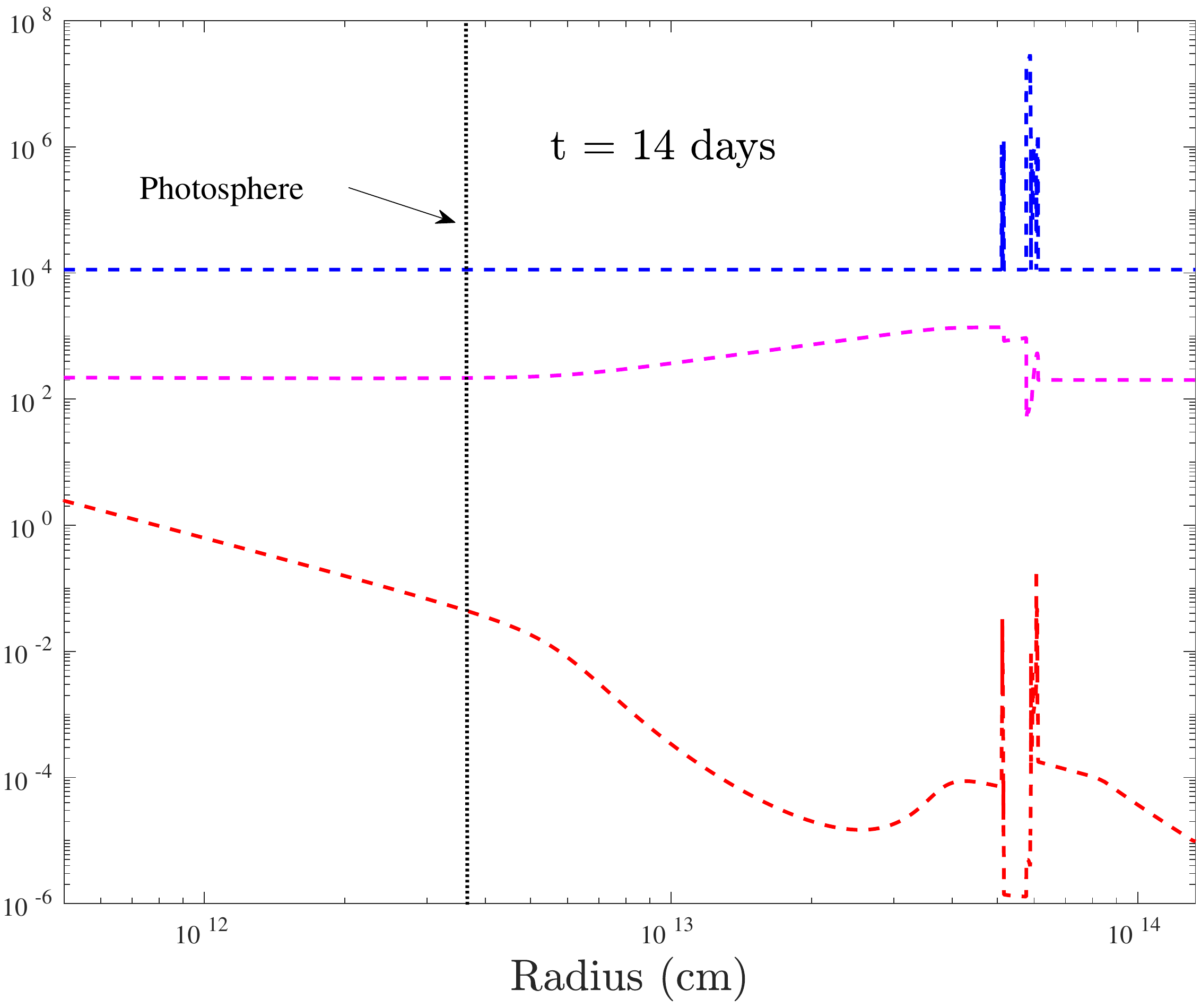}
    \caption{Snapshots of radial profiles of gas density, temperature, and radial velocity, for the multiple-transition model at $t$ = 8.5 days ({\it top}) and 14 days ({\it bottom}).  The temperature spikes and velocity discontinuities mark the locations of the shocks, while the density peaks show the thin shells of cool swept-up gas.  The approximate location of the optical photosphere is shown by a vertical dashed line in the top panel.  By the second snapshot, two of the three original shock structures have merged and all shocks have moved well above the photosphere.}
    \label{fig:snapshot}
\end{figure}

\begin{figure}
    \centering
    \includegraphics[width=0.5\textwidth]{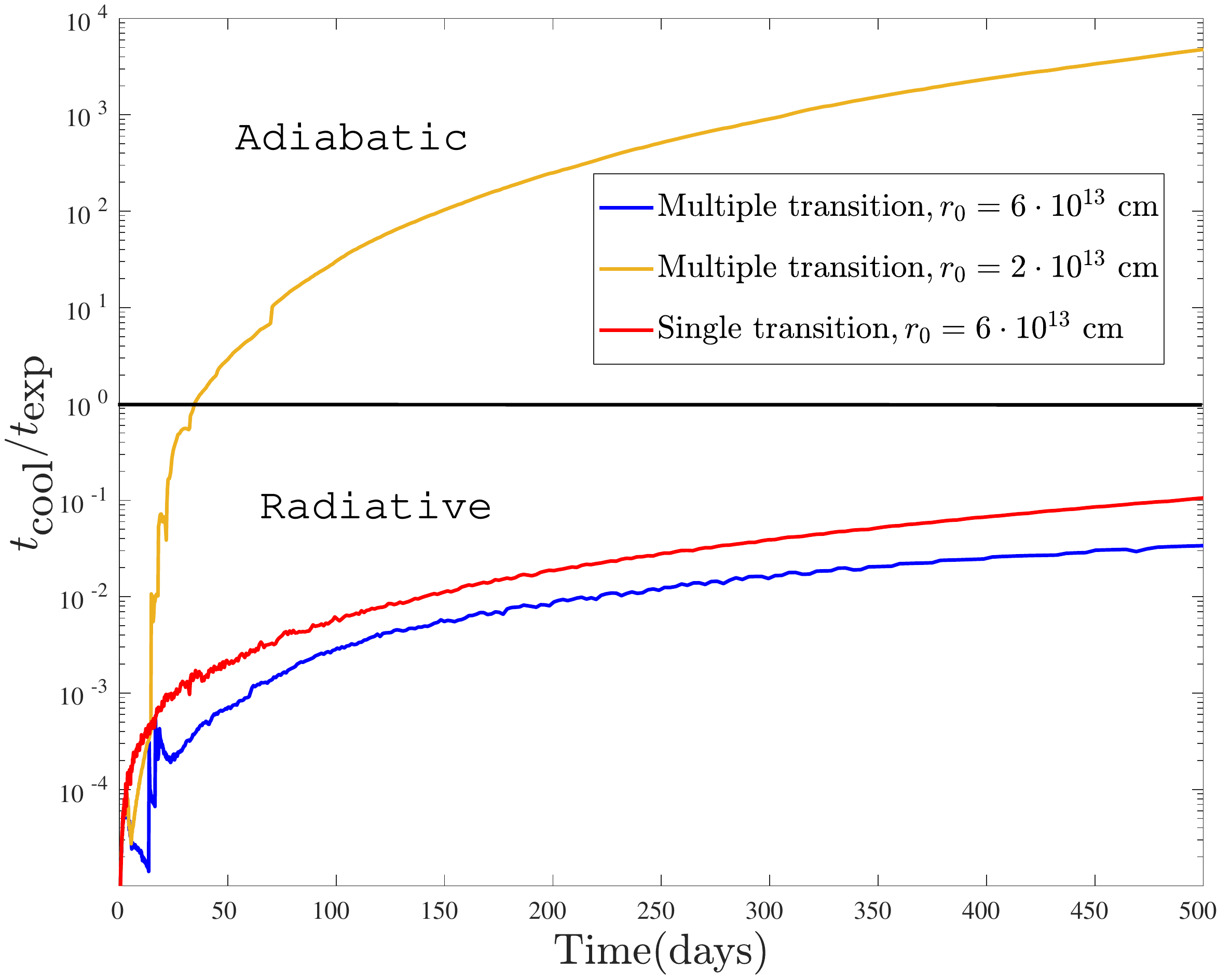}
    \caption{Ratio of the radiative cooling time of the immediate post-shock gas, $t_{\rm cool}$, to the radial expansion time, $t_{\rm exp} = r/v_{\rm sh}$, of the outermost forward shock as a function of time.  We show results separately for the single- and multiple-transition models.  All models shown here assume solar metallicity for the ejecta composition.}
    \label{fig:eta}
\end{figure}

By contrast, in the multiple transition model each new flip back to the fast outflow mode generates a new forward/reverse shock structure sandwiched around its own cool shell, which continues propagating out into the slow ejecta (see Figure \ref{fig:snapshot} for snapshots of the radial profiles of the ejecta properties).  These thin shells themselves collide at larger radii, creating a second (or third) generation of shocks, and ultimately merging pairwise into a single shell.  In all models, the final state is thus similar: a monolithic dense shell bounded in front by a forward shock propagating down the density gradient of the outermost layers of the  initial slow upstream medium.  Depending on the composition and density profile of the latter (bottom panel of Fig.~\ref{fig:inject}), the forward shock can remain radiative$-$with a radiative cooling time much shorter than that of radial expansion$-$for timescales of months or longer (Fig.~\ref{fig:eta}).

In all our models we arbitrarily shut off the injection after two weeks by having both the density and the velocity of the ejecta exponentially decay, the exact method of the decay does not influence our main results. The time scale of two weeks was chosen to give a peak luminosity time (T2) of order three weeks. Changing the onset of the decay would simply affect the T2 timescale.

\begin{figure}

  \includegraphics[width=\linewidth]{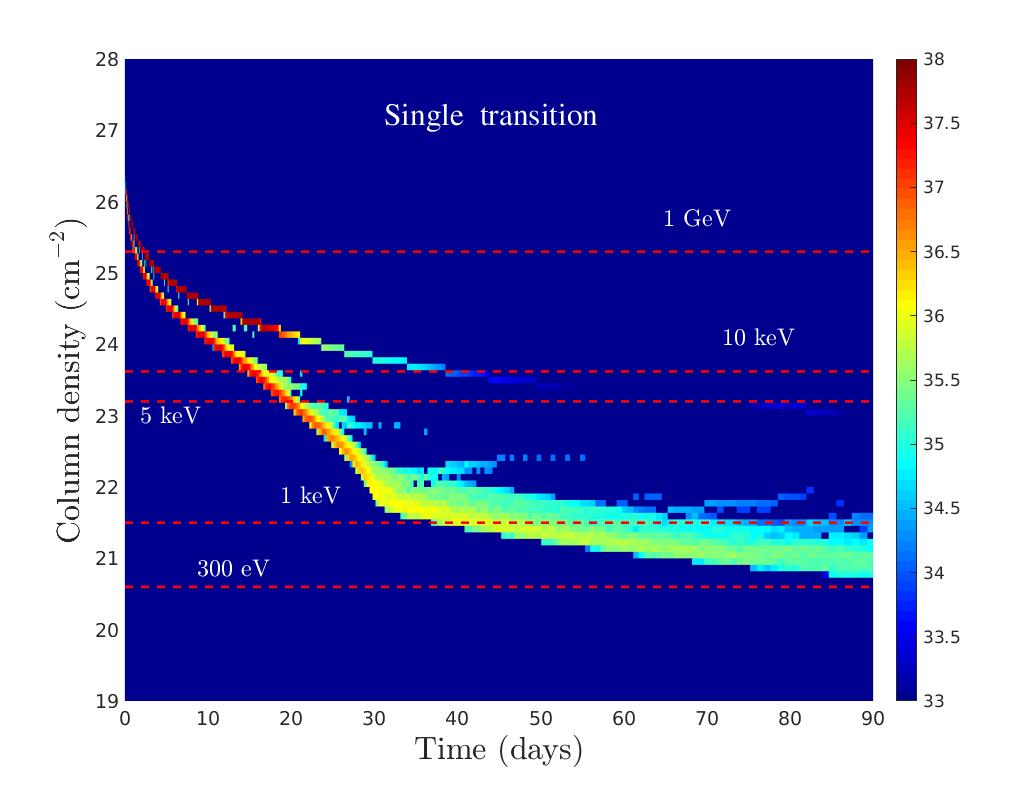}
  \includegraphics[width=\linewidth]{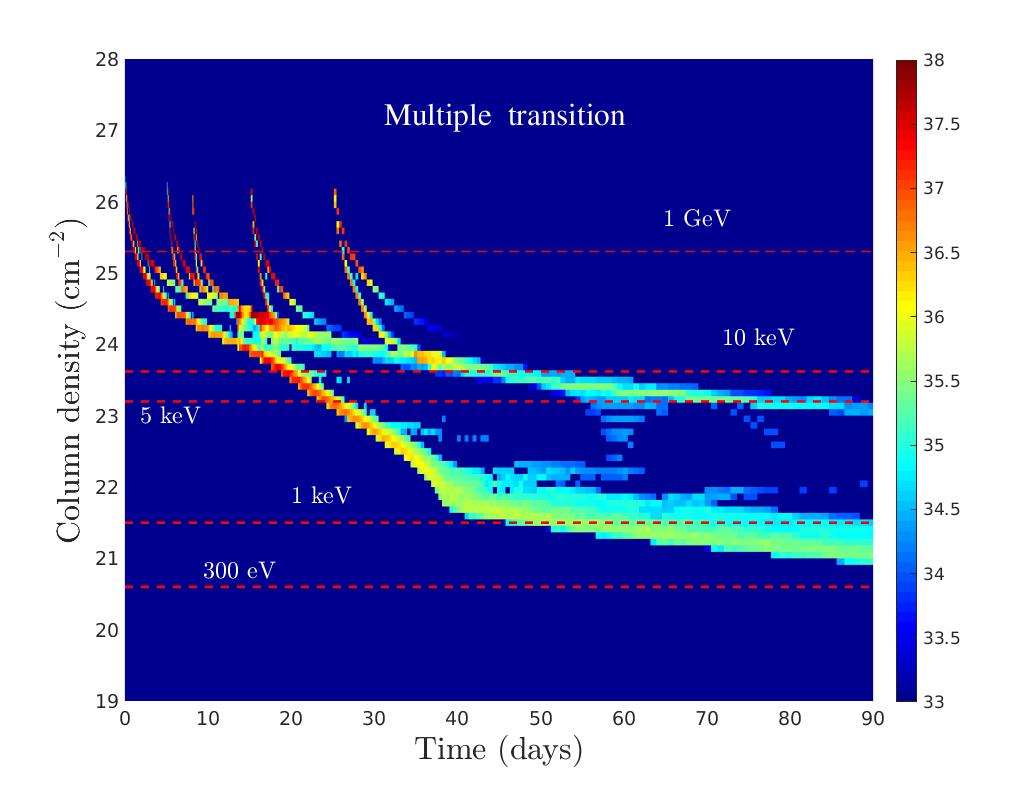}
  \caption{Radiated luminosity of the shock-heated gas as function of time and radial column depth ahead of the shocks, in units of erg s$^{-1}$/($\cdot$ cm$^{2}$), for the single-transition ({\it Top}) and multiple-transition ({\it Bottom}) models.  Horizontal dashed lines show the approximate columns below which photons of the different energies labeled emerge to the observer without significant absorption.}
  \label{fig:columns}
\end{figure}

Figure \ref{fig:columns} shows the radiated luminosity, which tracks the outwards movement of the shocks, as a function of the column density $\Sigma(r) = \int_{r}^{\infty} \rho dr'$ to the outflow surface, for the two fiducial models.  At early times, just after the onset of the fast flow, the shocks take place behind large columns $\Sigma \gtrsim 10^{22}-10^{25}$ cm$^{2}$, which are high enough to absorb the shock's UV/X-ray luminosity across a wide photon energy range $\sim 10$ eV $- 1$ keV for a month or longer after the outburst.  Following absorption and re-emission (which takes place effectively instantaneously), the shock's luminosity has been ``reprocessed'' into optical continuum or line emission.  The emission will emerge in the optical band because the ejecta opacity at optical wavelengths is much lower than in the UV/X-ray and the photon diffusion time is much shorter than the outflow expansion rate (such that losses due to adiabatic expansion can be neglected).  

\begin{figure}
    \centering
    \includegraphics[width=0.5\textwidth]{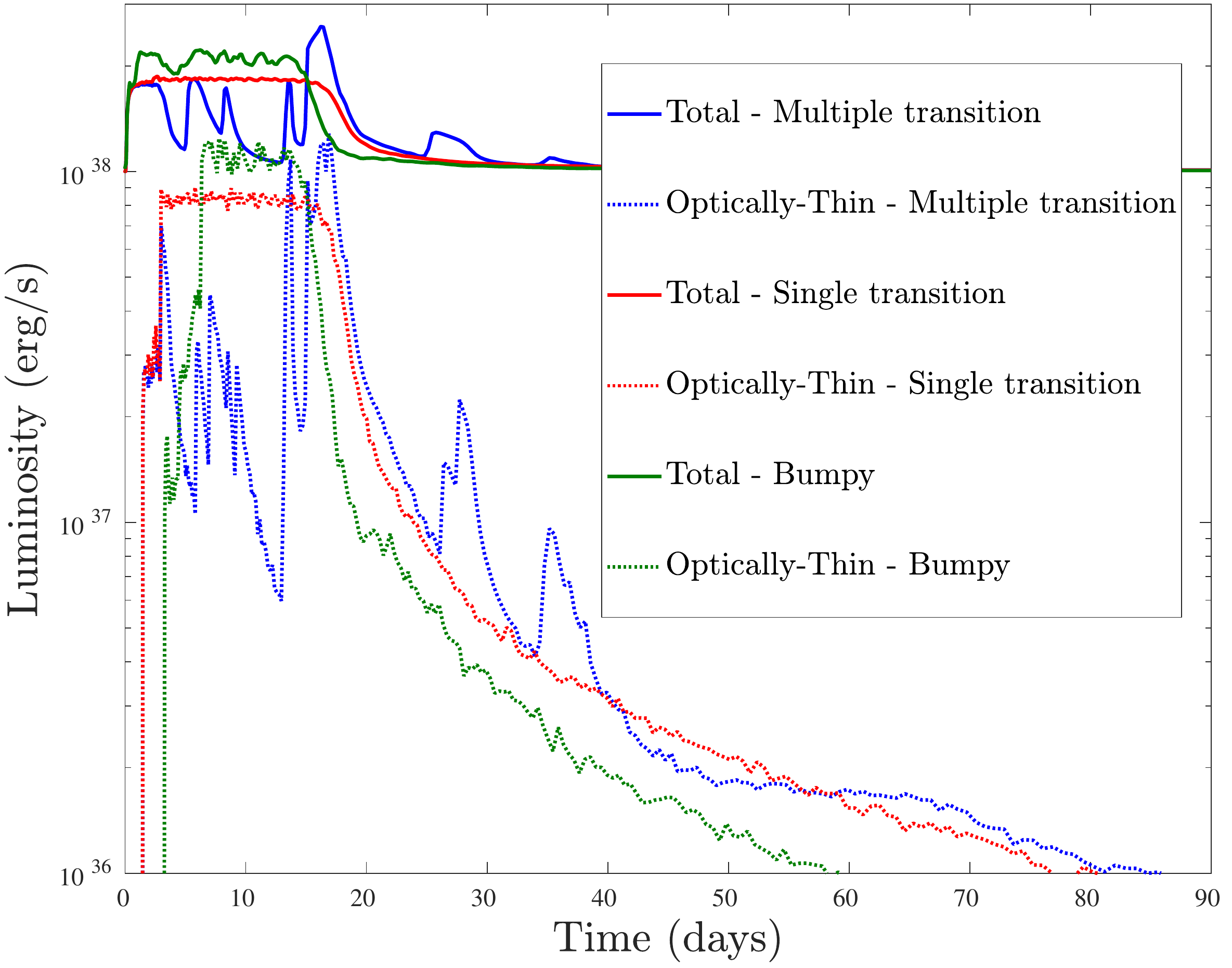}
    \caption{Solid lines show the total optical luminosity, $L_{\rm tot} = L_{\rm sh}(t) + L_{\rm wd}$, which includes contributions from the shocks $L_{\rm sh}$ and the central white dwarf $L_{\rm wd} = 10^{38}$ erg s$^{-1}$, which for simplicity we have taken to be constant in time.  Dotted lines show just the portion of the shock's luminosity $L_{\rm sh}$ emitted above the optical photosphere and which can therefore contribute to emission lines or optically-thin free-free emission.  We show results separately for the Fiducial single-transition and multiple-transition models, and the ``Bumpy" model in which the initial slow outflow is radially inhomogeneous. }
    \label{fig:Ltot}
\end{figure}

Under the reasonable assumption of instantaneous reprocessing, Figure \ref{fig:Ltot} shows the optical light curve of the  nova outburst (solid lines) in our different models.  The luminosity shown includes both the shocks as well as a contribution $L_{\rm wd} = 10^{38}$ erg s$^{-1}$ from the central white dwarf (also reprocessed by the ejecta at early times) and has been somewhat arbitrarily held constant in time.  Although the single transition scenario (red lines) produces a relatively smooth light curve, the multitude of collisions in the multiple-transition scenario (blue lines) generate several high amplitude flares.  That such a complex variable light curve can be generated, even from relatively simple assumptions about the outflow time-evolution and 1D geometry, illustrates that radiative shocks are a promising mechanism to generate flares in classical novae \citep{Strope+10}.  

Particularly bright and narrow flares occur in the multiple-transition model when two of the shock-generated thin shells collide and merge.  However, our calculations are likely overestimating the luminosity of such second-generation shocks for two reasons.  First, our 1D model does not account for angular spread in the outflow properties, i.e. in a realistic situation the shell collision will not take place simultaneously at all polar and azimuthal angles.  Second, our simulations are underestimating the thickness of the shells when they collide and thus the duration of the shock interaction.  Our 1D simulations do not capture the non-linear thin-shell instability \citep{Vishniac94,Steinberg&Metzger18}, which would act to corrugate the shell and thus increase its effective radial thickness.

\begin{figure}
    \centering
    \includegraphics[width=0.5\textwidth]{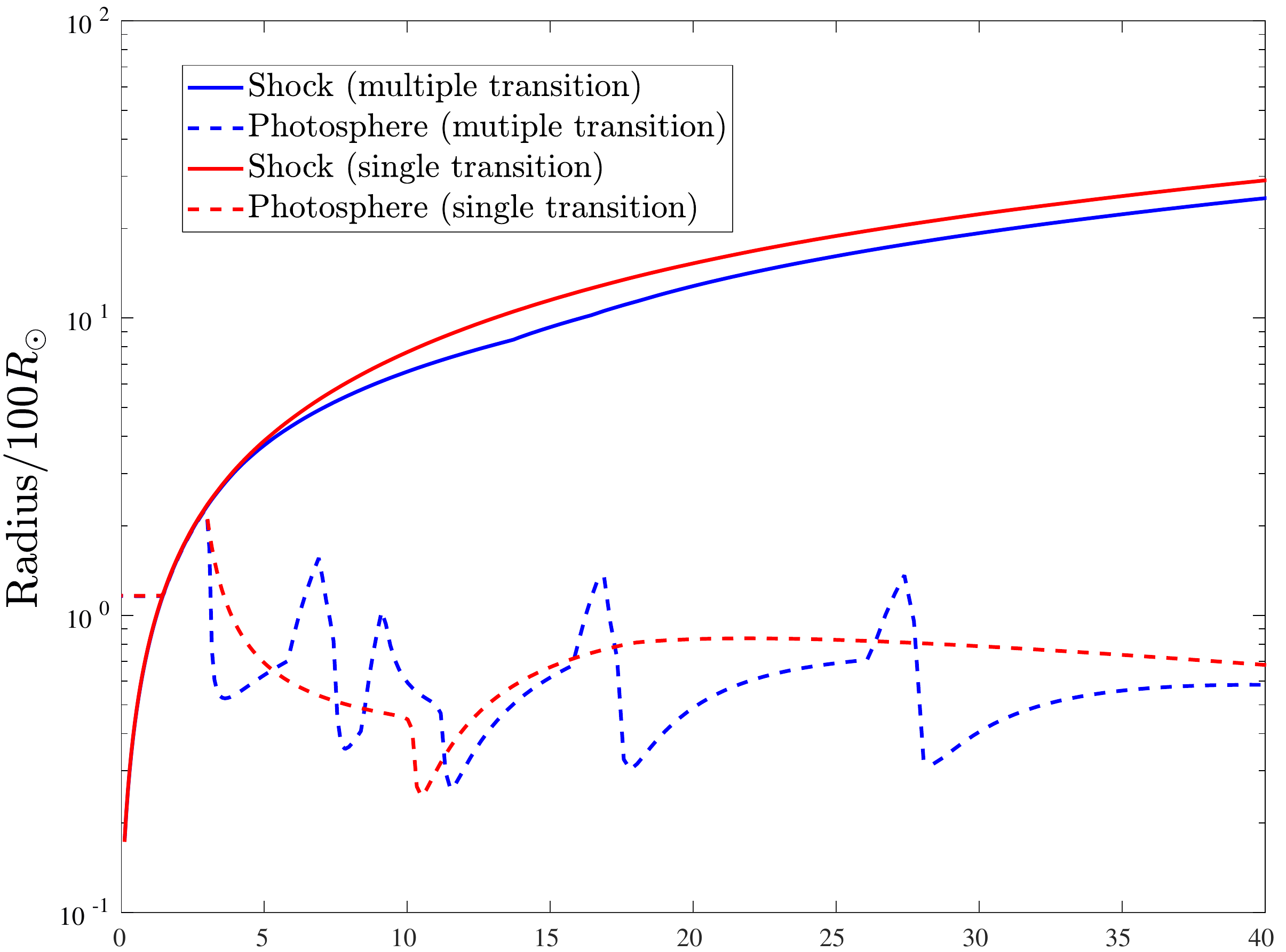}
 \includegraphics[width=0.5\textwidth]{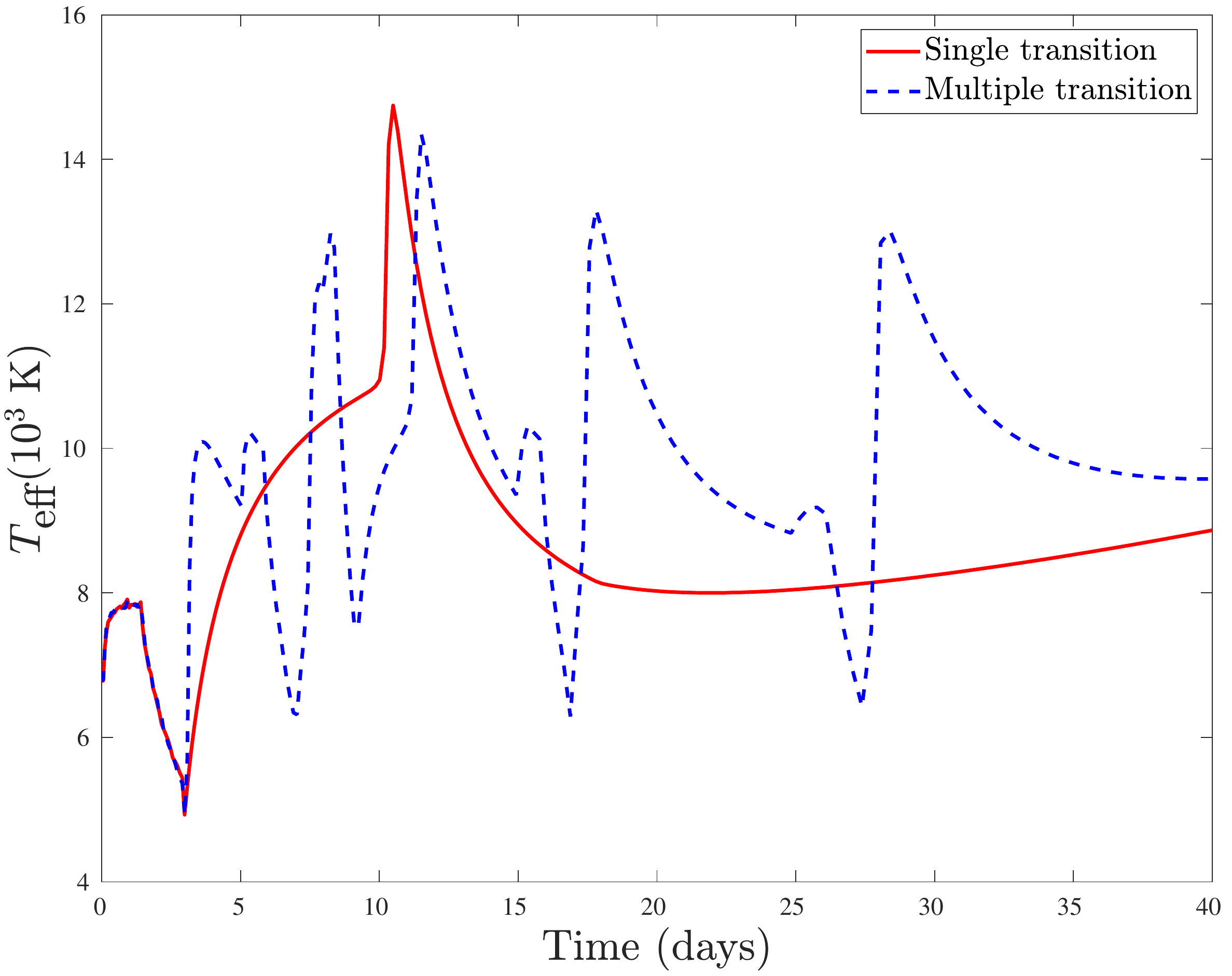}
    \caption{{\it Top:} Radius of the photosphere ({\it dashed lines}) compared to the radius of the outermost forward shock ({\it solid lines}).  {\it Bottom:} Effective temperature of the photosphere.}
    \label{fig:photo}
\end{figure}

We define the optical photosphere as the location where the radial optical depth obeys $\tau_{\rm opt} = \Sigma \kappa_{\rm opt} = 1$, where $\kappa_{\rm opt} = 0.03$ cm$^{2}$ g$^{-1}$ is an estimate of the effective opacity of the nova outflow to electron scattering and Doppler-broadened absorption lines (e.g.~\citealt{Pinto&Eastman00}).\footnote{Free-free opacity generally contributes negligibly at optical/UV wavelengths because the shocked gas rapidly cools to $\lesssim 10^{4}$ K, resulting in a moderate column of ionized gas.}  The dashed lines in Figure \ref{fig:Ltot} show just that portion of the shock luminosity which is reprocessed above the optical continuum photosphere, which we may therefore take as a rough proxy for the shock's contribution to emission lines and/or optically-thin free-free emission.  Figure~\ref{fig:photo} shows the radii of the outermost shock and the photosphere (top panel) as well as the blackbody temperature calculated from the radius and luminosity (bottom panel). 

In the single transition scenario, the shock quickly moves ahead of the photosphere.  Thus, after only a short delay, all of the shock-powered emission happens in optically-thin regions.  In such a scenario, in addition to  the simple light curve behavior, it would be challenging to explain the correlated behavior between the light curve and photosphere properties (see $\S\ref{sec:photo}$ below).  By contrast, in the multiple-transition scenario shell collisions can take place at small radii, near or beneath the optical photosphere, even well after the onset of the outburst.  Specifically, the bulk of the shock's emission will take place below the photosphere if the timescale over which the outflow undergoes the transition from slow-to-fast $\delta t$ is sufficiently short,
\begin{equation}
\begin{split}
    \delta t &\lesssim t_{\rm thin} \approx \frac{\dot{M}\kappa_{\rm opt}}{4\pi v_{\rm f}v_{\rm s}}
\\ \approx &1\,{\rm day}\,\left(\frac{\dot{M}}{10^{-4}\;M_\odot{\rm wk^{-1}}}\right)\left(\frac{1500\;\rm km\,s^{-1}}{v_{\rm f}}\right)\left(\frac{500\;\rm km\,s^{-1}}{v_{\rm s}}\right),
    \end{split}
\label{eq:t_thin}
\end{equation}
where $t_{\rm thin}$ is the time required for a thin shell of velocity $\sim v_{\rm f}$ takes to reach the photosphere of the slow upstream wind of mass-loss rate $\dot{M}$ and velocity $v_{\rm s}$.  Depending on the precise values of $\dot{M}$ and $\delta t$, it is possible to get shocks that release most of their energy below, or above, the photosphere.

Green lines in Figure \ref{fig:Ltot} show the predicted light curves of the ``Bumpy" model in which (as in the single transition model) the outflow velocity evolution is monotonic, but rather than a steady-wind the initial slow outflow is radially inhomogeneous (bottom panel, Fig.~\ref{fig:inject}).  In this case, one obtains a multi-peaked light curve, as the single shock structure generated by the fast transition propagates into the external-most slow medium.  However, the amplitude of the flares are modest (tens of percent) compared to the larger variations observed in some novae and those seen in the multiple-transition model.  Furthermore, the shock quickly moves ahead of the photosphere, which (as in the single-transition model) precludes again correlated behavior in light curve and photosphere properties.  Variations in the $\dot{M}$ of the initial slow outflow alone do not appear sufficient to explain flares similar to those observed in novae.


\subsection{Photosphere Evolution During Flares}
\label{sec:photo}

Time-resolved optical spectroscopy show that new absorption line systems appear in novae at times coinciding with light curve flares (e.g.~\citealt{Jack+17}).  \citet{Tanaka+11a,Tanaka+11b} argue that such appearances/disappearances of absorption features are caused by the temporary expansion/shrinking of the optical photosphere.  During maxima, the optical photosphere expands in size and the absorption components appear as the ejecta shell expands to become optically-thin.  During mimina, the optical photosphere shrinks and the absorption component weakens relative to the emission lines.  As we now discuss, such behavior can arise naturally from the multiple-transition internal shock scenario.

When shocks start below the photosphere, this causes an almost instantaneous rise in the optical luminosity, $L_{\rm tot}$.  This is because the timescale for photon diffusion from the shock through the ejecta, $\tau_{\rm opt}(R_{\rm sh}/c)$, is short compared to the outflow expansion timescale, $R_{\rm sh}/v$, provided that $\tau_{\rm opt} \lesssim c/v \sim 10^{2}-10^{3}$ ($\Sigma \lesssim 10^{25}-10^{26}$ cm$^{-2}$), where $c$ is the speed of light and $v$ is the radial velocity of the emitting gas.  The photosphere radius $R_{\rm ph}$ is initially unchanged before the shock reaches it.  Therefore, the new contribution to the luminosity from the shocks causes an immediate increase in the effective temperature of the continuum,
\be
T_{\rm eff} = \left(\frac{L_{\rm tot}}{4\pi\sigma R_{\rm ph}^{2}}\right)^{1/4}. 
\ee 
After a delay $\sim t_{\rm thin}$ (eq.~\ref{eq:t_thin}) the shock emerges from the photosphere of the slow outflow; the dense shell behind it carries the photosphere temporarily outwards with it, causing $R_{\rm ph}$ to rise and thus $T_{\rm eff}$ to {\it decrease}.  A time lag $\Delta t$ therefore separates maxima in $T_{\rm eff}$ and $R_{\rm ph}$, which is approximately given by the timescale $t_{\rm thin}$ required for the swept-up shell to reach the photosphere (eq.~\ref{eq:t_thin}).  

\begin{figure}
  \includegraphics[width=\linewidth]{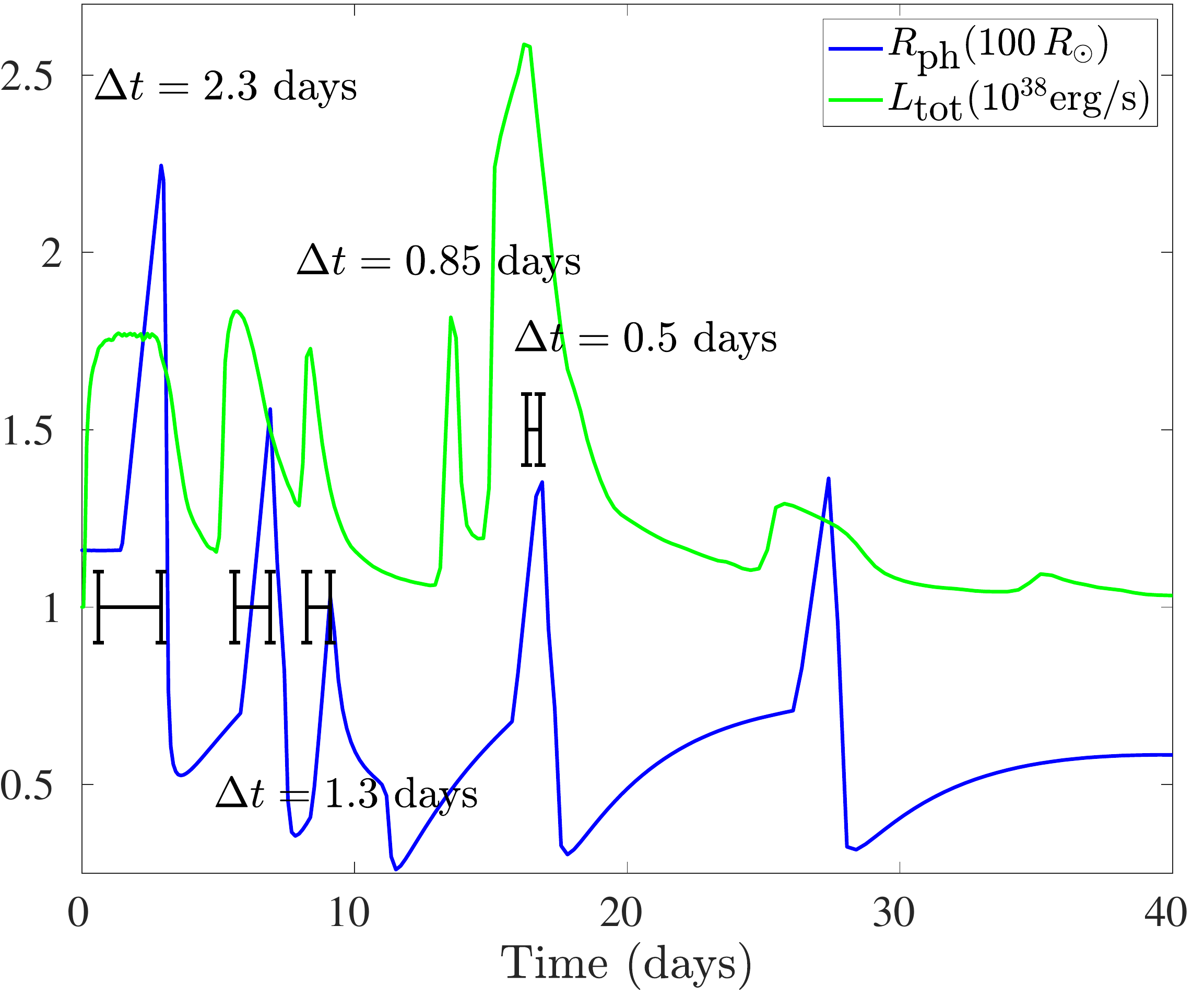}
  \caption{Correlated evolution (with a modest time lag) of the total luminosity $L_{\rm tot}$ ({\it blue}) and photosphere radius $R_{\rm ph}$ ({\it green}) in the multiple-transition model.  Because each shock starts beneath the photosphere, the maxima in $L_{\rm tot}$ (which effectively instantaneously tracks the shock power) precede maxima in $R_{\rm ph}$ (which occur only once the cool shell swept up by the shocks reaches the photosphere and temporarily carries it outwards).}
  \label{fig:LR}
\end{figure}

Figure \ref{fig:LR} provides a side-by-side comparison of the evolution of $L_{\rm tot}$ and $R_{\rm ph}$ for the multiple-transition model.  Each maxima in $L_{\rm tot}$ and $R_{\rm ph}$ corresponds to the emergence of a given shock-bounded shell, for which we find delays in the range $\Delta t \approx 0.5-2.3$ days.  Such anti-correlated $L_{\rm tot}/R_{\rm ph}$ behavior with a delay is seen in novae such as ASASSN-17pf (\citealt{Aydi+19}), though the lag may in some cases be too short to observe.

\begin{figure}
  \includegraphics[width=\linewidth]{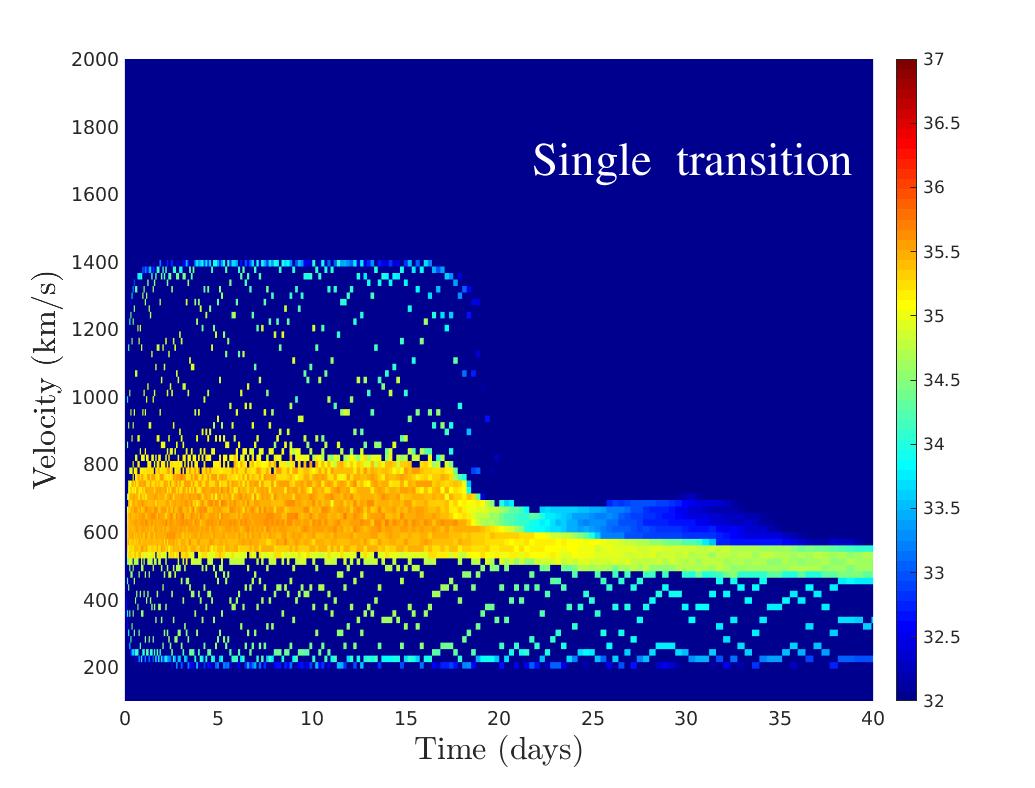}
  \includegraphics[width=\linewidth]{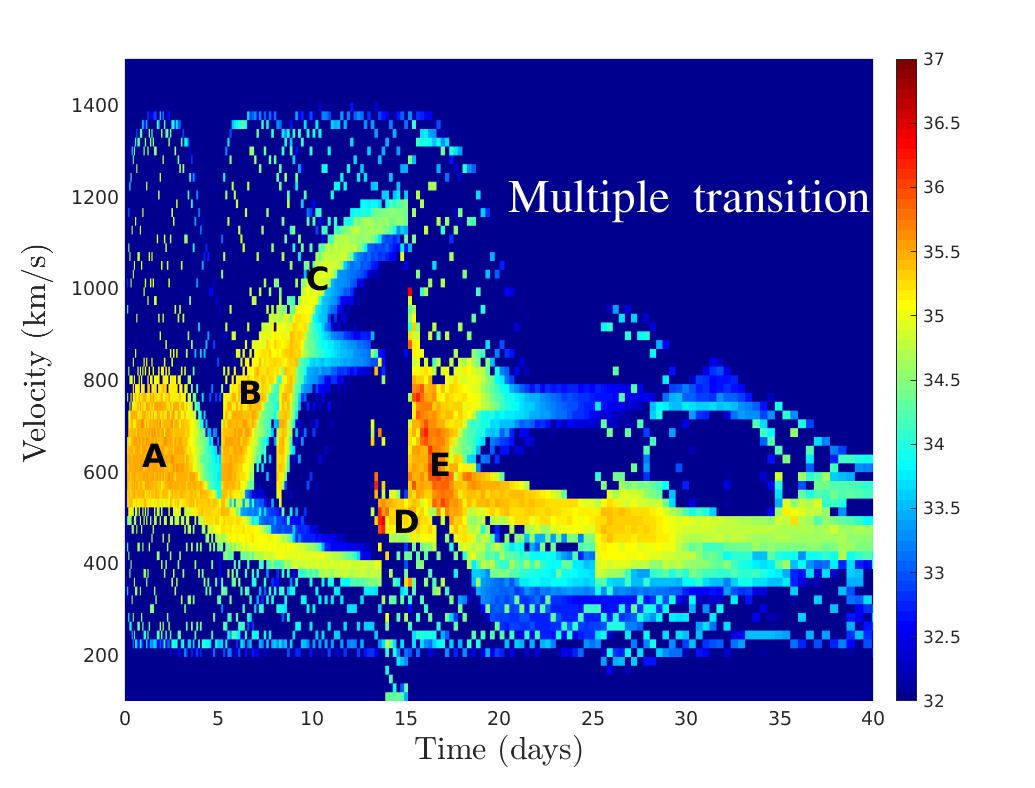}
  \caption{Shock luminosity as a function of time, now broken down by the radial velocity of the emitting gas, for the single-transition ({\it top}) and multiple-transition ({\it bottom}) models.  The units on the color scale are erg s$^{-1}$/(km s$^{-1}$).}
  \label{fig:dL_dv}
\end{figure}

Figure \ref{fig:dL_dv} shows the time evolution of shock luminosity in the single and multiple-transition models, but now broken down by the radial velocity of the emitting gas.  For shocks which take place near or above the continuum photosphere, these provide a rough proxy for the Doppler widths of features in the optical spectrum (e.g. emission lines or P-Cygni line profiles).   However, note that in detail the observed line widths will differ from those predicted by Fig.~\ref{fig:dL_dv}, as our 1D models do not account for angular spread and line-of-sight projection effects in the outflow.  We are justified in assuming that the optical emission occurs locally near the cooling gas in our simulation, because of the very short mean free path over which the UV and soft X-ray photons (which carry most of the shock's power; \citealt{Steinberg&Metzger18}) are reprocessed.   

The single-transition model (top panel of Fig.~\ref{fig:dL_dv}) predicts smooth time evolution of line features,  covering a range of velocities 500$-$800 km s$^{-1}$ centered around that of the cool shell.  The emission is dominated by the layers of cooling gas behind the forward and reverse shock, which are geometrically narrow but span a range of speeds as the emitting gas settles sub-sonically onto the central cool shell.  

In contrast to the relatively simple evolution of the single-transition model, the multiple-transition model (bottom panel of Fig.~\ref{fig:dL_dv}) shows multiple complex velocity features, including both decelerating and {\it accelerating} components.  The first decelerating feature in Fig.~\ref{fig:dL_dv} (marked A) is emission from the forward/reverse shock structure generated by the first fast outflow; the fast outflow turns off after a few days, but the forward shock from this shell continues to dominate the luminosity as it sweeps up mass and decelerates.  The shocks generated from the second (B) and third (C) transitions generate accelerating line features as the emitting gas is being pushed by their respective fast outflows.  Eventually (D), the front end of the shell generated by transitions B catches up to the rear of the shell produced by the first transition (A), leading to a collision between the shell and a brief flare (which as we discussed earlier, may not be accurately captured by a 1D model due to the artificially thin shells).  Around  the same time, the fourth transition (E) occurs as well as a collision between the second and third shells (B/C).

Acceleration, deceleration and even the apparent ``merger'' of velocity components are common features of time-resolved spectra of classical novae (e.g.~\citealt{Walter+12,Aydi+19}).  Although detailed radiative transfer on our simulation data would be required to make more definitive statements, our findings suggest that internal radiative shocks, near or above the optical photosphere, provide a promising explanation for this observed behavior.

\subsection{Gamma-Ray Emission}
\label{sec:gammarays}

\begin{figure}
    \centering
 \includegraphics[width=0.5\textwidth]{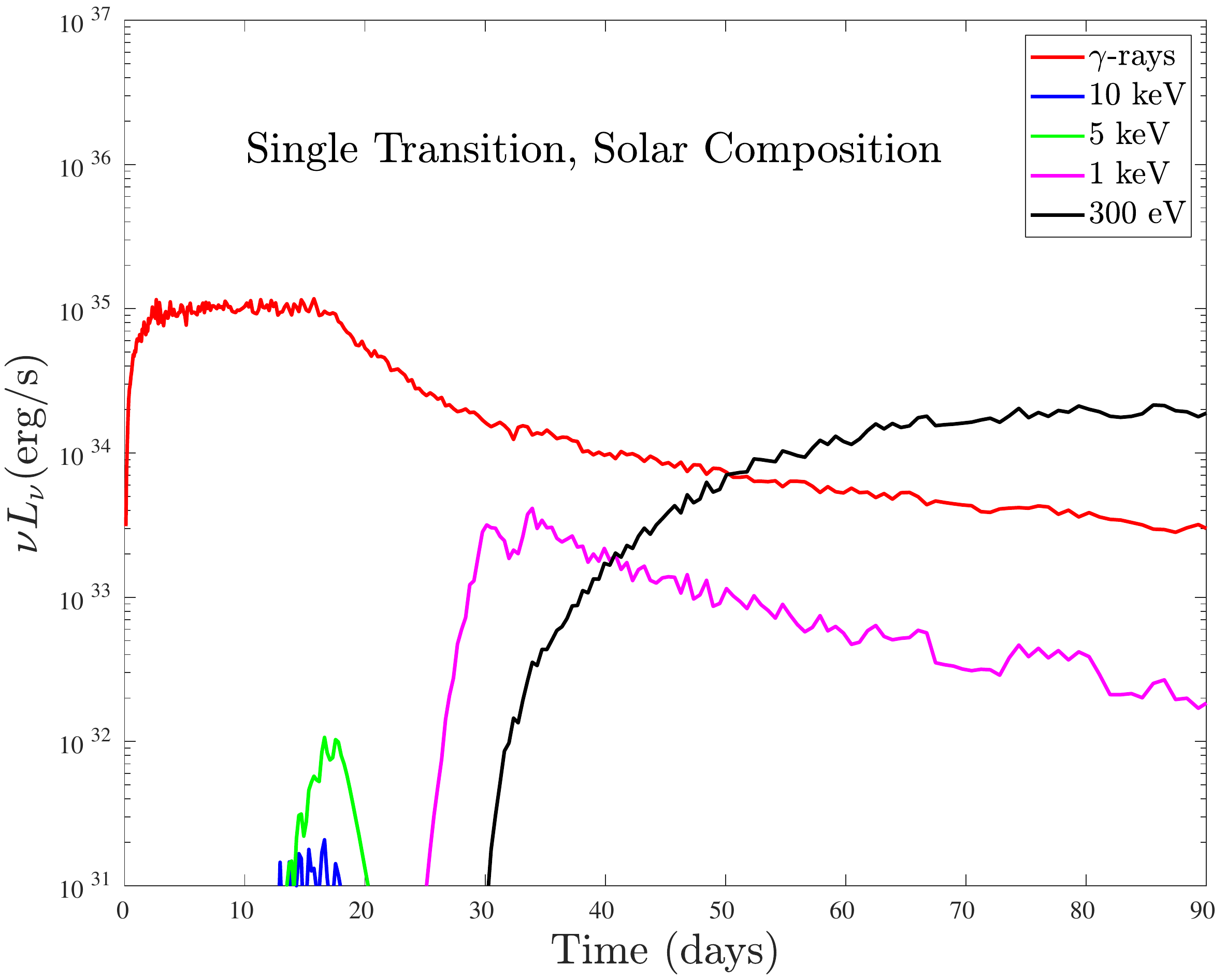}
    \includegraphics[width=0.5\textwidth]{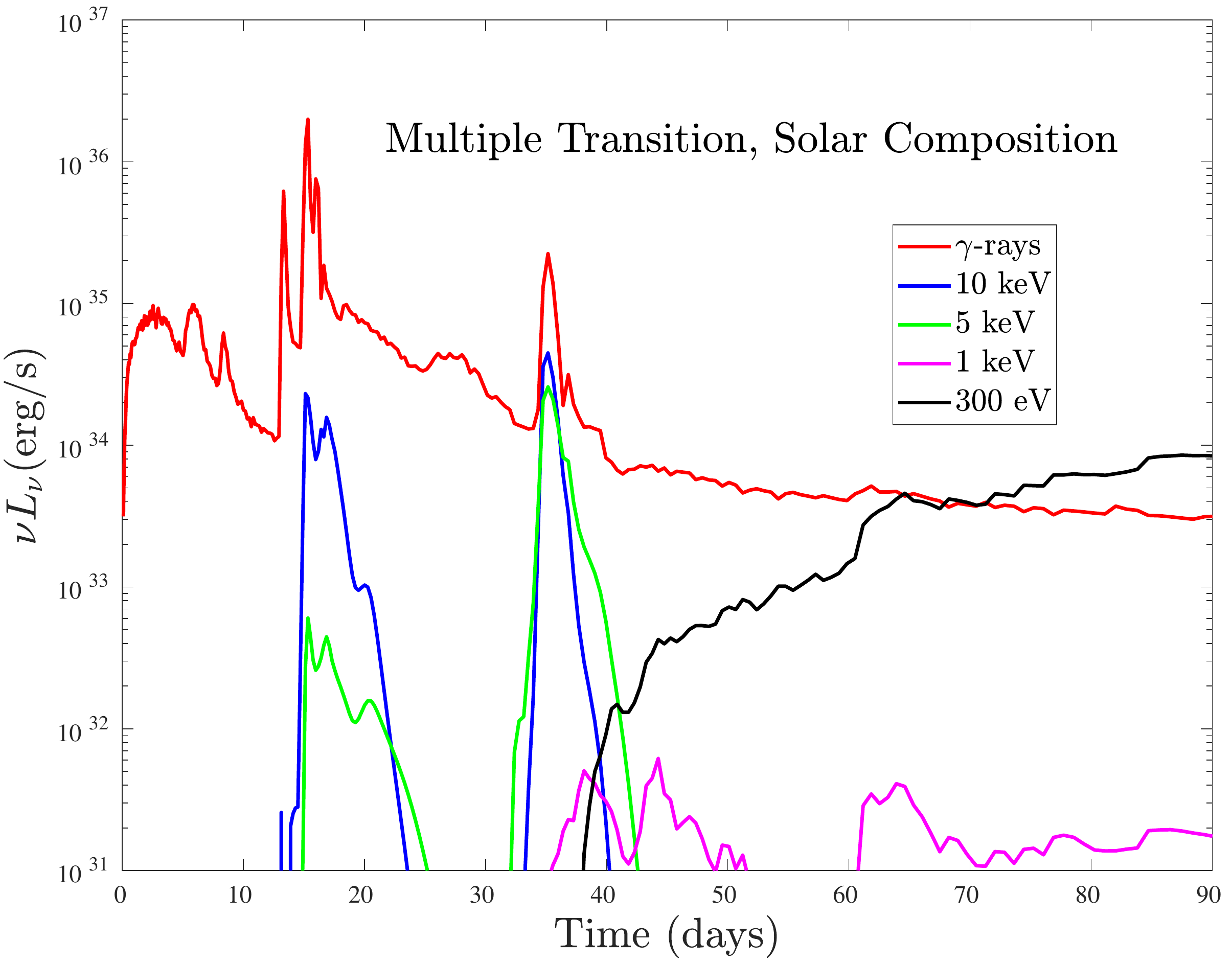}
    \caption{Total shock-powered gamma-ray luminosity and $\nu L_{\nu}$ thermal X-ray luminosity (at various photon energies as labeled by the line color) in the single-transition ({\it Top}) and multiple-transition ({\it Bottom}) models.  Here we assume solar abundances in calculating the gas cooling and ejecta opacity.}
    \label{fig:xray}
\end{figure}

\begin{figure}
    \centering
 \includegraphics[width=0.5\textwidth]{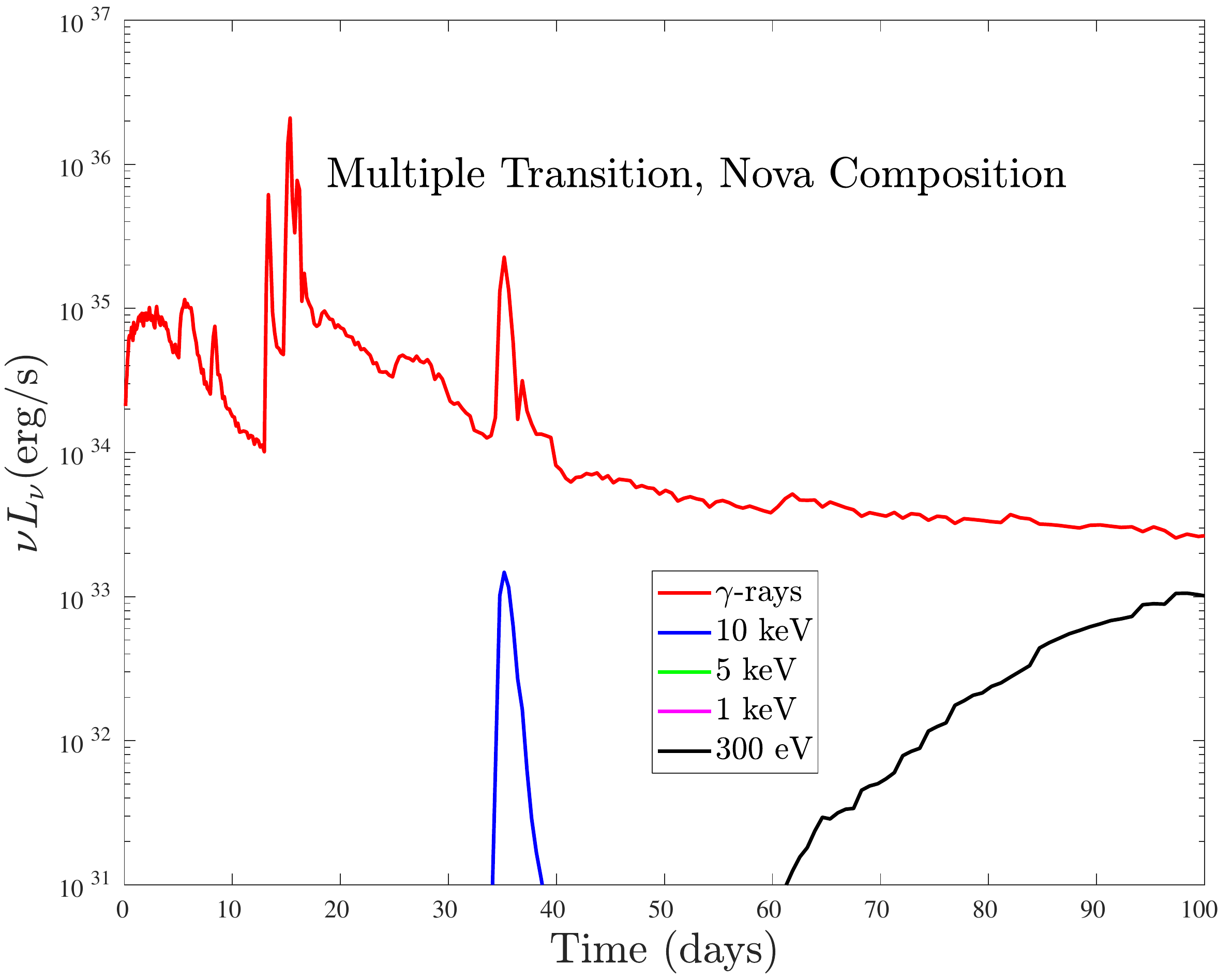}
    \includegraphics[width=0.5\textwidth]{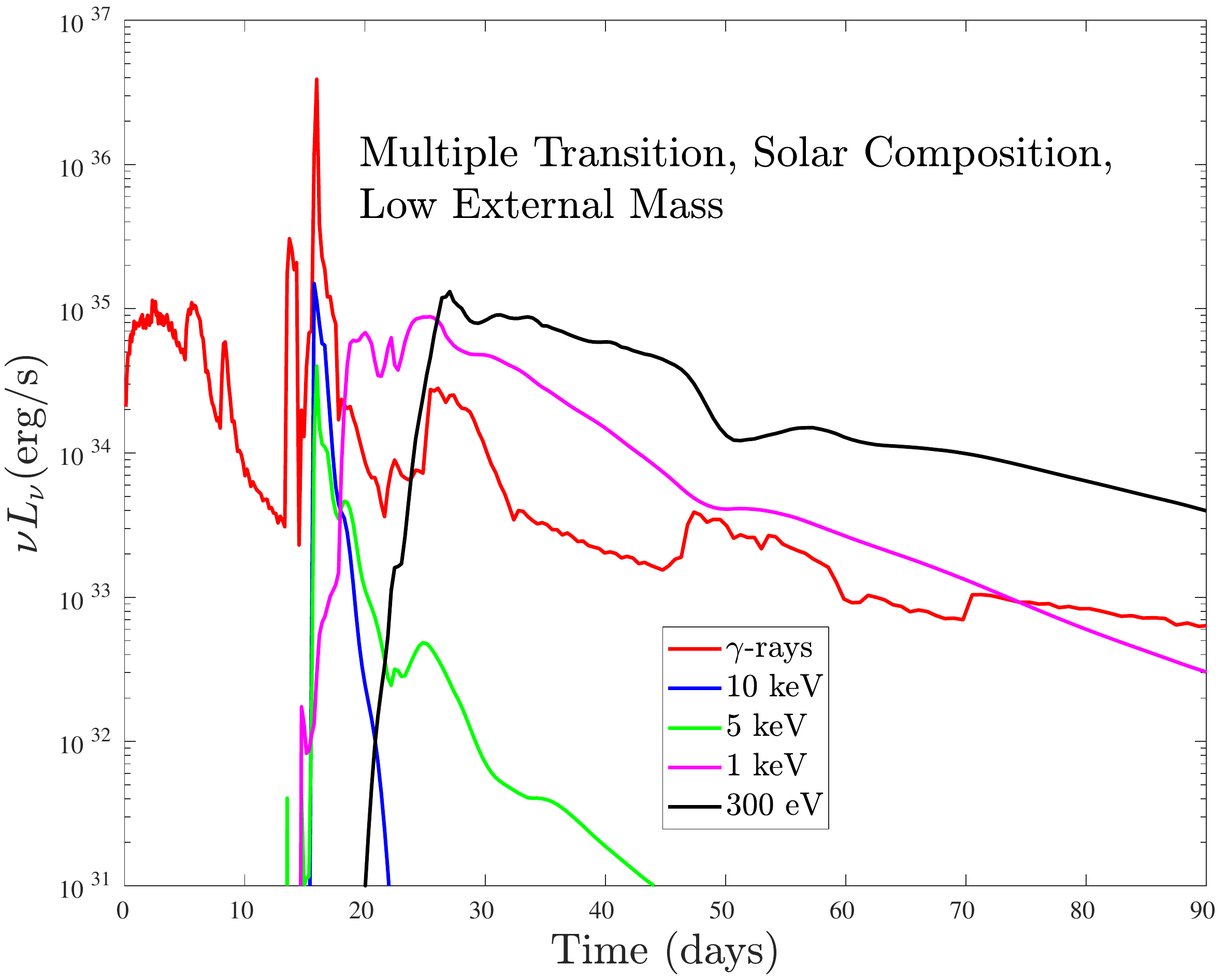}
    \caption{Same as Fig.~\ref{fig:xray}, but showing the effect of assuming CNO-enhanced {\it nova} abundances as defined in CLOUDY instead of solar composition ({\it top}) or a lower mass ($r_0 = 2\times 10^{13}$ cm) for the initial slow medium ({\it bottom}).}
    \label{fig:xray2}
\end{figure}

Figures \ref{fig:xray} and \ref{fig:xray2} show the predicted X-ray and gamma-ray light curves from our models.  The $\sim$ GeV gamma-rays detected from novae by {\it Fermi} LAT \citep{Ackermann+14} are likely produced either by (a) relativistic ions accelerated at the shock, which collide in-elastically with ambient ions to generate neutral pions that quickly decay into gamma-rays; (b) relativistic electrons accelerated at the shock (or secondary electron/positron pairs produced from charged pion decay), which in-elastically scatter ambient nova optical photons up into the gamma-ray band via the Inverse Compton process, or which interact with ambient gas and emit gamma-rays through the (relativistic) bremsstrahlung process.  In either of these (a) ``hadronic'' or (b) ``leptonic'' models, the timescale for relativistic particles to lose their energy and generate gamma-rays is generally short compared to the outflow expansion timescale over which adiabatic losses occur \citep{Metzger+16,Martin+18,Vurm&Metzger18}.  

The main source of opacity for gamma-rays in the $\sim 0.1-10$ GeV energy range is due to photon-matter pair creation, for which $\kappa_{\gamma} \approx 0.004$ cm$^{2}$ g$^{-1}$ (e.g.~\citealt{Zdziarski&Svensson89}).  Figure \ref{fig:columns} shows that, in both single- and multiple-transition models, the shocks spend most of the outburst duration above the gamma-ray photosphere.  
Nevertheless, the shocks can be deeply embedded at early times after the fast outflow is launched ($\tau_{\gamma} = \Sigma \kappa_{\gamma} >1$; $\Sigma \gtrsim 10^{26}$ cm$^{2}$ g$^{-1}$) and this could generate a modest delay in the onset of gamma-ray emission relative to the optical peak of a flare \citep{Li+17}.  

Given these two facts -- fast-cooling cosmic rays and gamma-ray-transparent ejecta -- we expect the observed gamma-ray emission to approximately track the instantaneous power of the shock, much as with the reprocessed optical emission.  Indeed, correlated gamma-ray and optical light curves are seen in both cases with time-resolved gamma-ray data: ASASSN-15ma \citep{Li+17} and ASASSN-18fv (Aydi et al., in prep).  \citet{Metzger+15} show how this behavior allows for an estimate of the efficiency of relativistic particle acceleration at the shocks, $\epsilon_{\gamma} \sim L_{\gamma}/L_{\rm sh}$, where $L_{\gamma}$ and $L_{\rm sh}$ are, respectively, the gamma-ray luminosity and the portion of the optical luminosity powered by shocks.  Application of this method to gamma-ray detected novae results in typical values of $\epsilon_{\gamma} \approx 3-5\times 10^{-3}$ (\citealt{Li+17}), consistent with theoretical expectations for the efficiency of diffusive shock acceleration.\footnote{Once one takes into account the average inclination angle of the upstream magnetic field relative to the shock front, taking into account its highly-corrugated shape when the shock is radiative \citep{Steinberg&Metzger18}.}  The gamma-ray light curves in Figs.~\ref{fig:xray}, \ref{fig:xray2} were calculated by multiplying the kinetic power of the shocks from the simulation by a constant value $\epsilon_{\gamma} = 3\times 10^{-3}$.  Our fiducial models predict peak gamma-ray luminosities of $\sim 10^{35}-10^{36}$ erg s$^{-1}$, which are consistent with the range observed by {\it Fermi} LAT \citep{Ackermann+14,Linford+15,Cheung+16}.  Our results are also consistent with \citet{Martin+18}, who show that the gamma-ray properties of LAT-detected novae can be understood as hadronic interactions following particle acceleration in internal shocks between faster and slower outflows (the analog of our single-transition model).

\subsection{Thermal X-Ray Emission}
\label{sec:Xrays}

In contrast to the non-thermal gamma-rays, the dominant source of X-ray emission from the shocks is thermal free-free emission.  Non-thermal X-rays (e.g. from relativistic bremstrahlung or inverse Compton scattering emission) are generally suppressed because the electrons that emit in the hard X-ray band possess relatively low energies compared to those emitting in the gamma-ray band and are subject to severe losses to Coulomb scattering off the dense background of the nova ejecta \citep{Vurm&Metzger18}.  Also in contrast to gamma-rays, the X-ray emission is significantly attenuated by photo-electric absorption in the ejecta.  The X-ray light curves shown in Fig.~\ref{fig:xray} and \ref{fig:xray2} are calculated assuming each cell $i$ at radius $r_i$ in the simulation emits free-free emission at its local temperature $T_i$, accounting for photoelectric attenuation according to $e^{-\tau_{\rm X,i}}$, where $\tau_{\rm X,i} = \int_{r_i}^{\infty} \kappa_{\rm X}\rho dr$ and $\kappa_{\rm X}$ is the opacity at the X-ray energy of interest. 

The X-ray light curves in Figure \ref{fig:xray} are calculated assuming solar composition for the ejecta opacity.  In our multiple-transition model, the gamma-ray/optical flares are accompanied by hard X-ray $\gtrsim 5-10$ keV flares of luminosity $L_{\rm X} \sim 10^{33}-10^{34}$ erg s$^{-1}$.  At face value, these luminosities appear to be broadly consistent with the {\it NuSTAR} detection of thermal X-ray emission contemporaneous with the {\it LAT} gamma-ray detection of V5855 Sgr (\citealt{Nelson+19}).  However, \citet{Nelson+19} infer very high temperatures for the un-absorbed X-ray emitting-plasma ($kT \sim 5-20$ keV) which require a higher shock velocity $\gtrsim 2000-4000$ km s$^{-1}$ than is obtained in our fiducial model.  If we had instead adopted a much faster white dwarf outflow velocity $v_{\rm f}$ (for otherwise similar kinetic energy), our predicted {\it NuSTAR} band X-ray fluxes would be several orders of magnitude greater $\gtrsim 10^{36}-10^{37}$ erg s$^{-1}$ than those shown in Fig.~\ref{fig:xray}, too high compared to observations.  

As discussed by \citet{Nelson+19}, the X-rays observed from V5855 Sgr are thus in fact {\it sub-luminous} relative to expectations given the high observed gamma-ray luminosity.  To reconcile this fact, one is forced to conclude that either the X-rays component which dominates in {\it NuSTAR} originates from distinct shocks from those which are responsible for powering the gamma-rays, i.e.~if the latter are too deeply buried for even hard X-rays to escape.  Alternatively, the hard X-ray luminosity of the gamma-ray-producing shocks could be suppressed from the naive predictions of a 1D model.  Supporting the latter possibility, \citet{Steinberg&Metzger18} found that the X-ray luminosity of high-Mach number $\mathcal{M} \gg 1$ radiative shocks is suppressed by a factor of $\sim 4.5/\mathcal{M}^{2} \sim 1/100$ for radiative shocks relative to the 1D predictions (see also \citealt{Kee+14}).  As mentioned above, a non-thermal origin for the hard X-rays from V5855 (e.g. bremsstrahlung emission from the relativistic particles generating the gamma-rays) is disfavored, due to both the luminosity and spectral slope of the emission \citep{Vurm&Metzger18}.  

In contrast to $\gtrsim 10$ keV X-rays, lower-energy X-rays (e.g. in the $\lesssim 5$ keV {\it Swift} XRT bandpass) are suppressed for several months or longer after the eruption.  The $\sim$ keV X-ray light curves in our models peak only once the forward shock of the outermost monolithic shell (created by the merger of previous shells) reaches a sufficiently low external column $\tau_{X} \lesssim 1$.  Our models predict peak keV luminosities from the shocks which range from undetectably weak to $L_{\rm X} \sim 10^{34}-10^{35}$ erg s$^{-1}$, depending sensitively on the mass, radial distribution, and composition of the slowly expanding external medium which exists prior to the onset of the first fast outflow.  

Figure \ref{fig:xray2} shows how the predicted X-ray light curves depend on the properties of the pre-existing external slow medium, for an otherwise identical outflow behavior at $t > 0$ (after the onset of the first fast outflow).  Specifically, we consider how the results change if we adopt the standard CNO-enhanced {\it nova} abundance set from CLOUDY \citep{Ferland+17} instead of solar abundances.  We also consider results for the Low Mass model, in which the pre-existing slow external medium is three times less massive than in the Fiducial model due to its smaller initial radial extent $r_0 = 2\times 10^{13}$ cm (Fig.~\ref{fig:inject}).  The greater keV opacity of enhanced CNO abundances compared to the solar metallicity case reduces the column $\Sigma$ ahead of the shock at the time of the X-ray peak ($\tau_{X} = 1$), thus resulting in a lower shock power $\propto \Sigma v_{\rm sh}^{3}$ and correspondingly lower X-ray luminosity.  The effect of smaller $r_0$ and lower mass in the external medium is to increase the forward shock velocity $v_{\rm sh}$ (since the shell has less mass to sweep up) and shorten the timescale over to achieve $\tau_X = 1$, both which increase the peak X-ray luminosity.  The wide range of X-ray luminosities our models find, even for moderate variations in the nova outflow properties, appears to be at least qualitatively consistent with the diverse range of X-ray light curve behavior in novae \citep{Mukai&Ishida01,Orio+01a,Orio+01b,Mukai+08,Schwarz+11}.  

The late-time onset of soft X-rays in classical novae is commonly interpreted as emission from the white dwarf surface (e.g.~\citealt{Schwarz+11}).  However, we note that in some cases such as our Fiducial model with a high external mass, the forward shock velocity is sufficiently low at late times for the X-ray emission to peak at low energies $\lesssim$ 0.3 keV (bottom panel of Fig.~\ref{fig:xray}) in which case shock emergence could masquerade as the onset of the supersoft phase.  On the other hand, differences in the predicted spectral energy distribution (blackbody emission from the white dwarf surface versus optically-thin emission from shocks at much larger radii) should be discernible with good spectral coverage.  

Also note that our 1D models do not take into account the suppression of X-ray emission from radiative shocks found by \citet{Steinberg&Metzger18} and therefore the luminosities shown in Figs.~\ref{fig:xray}, \ref{fig:xray2} may be over-estimates.  This caveat becomes less severe with time, as the forward shock transitions from being radiative to adiabatic over a month or longer (depending on the velocity of the shocks and the composition of the upstream medium; Fig.~\ref{fig:eta}).  

\subsection{Radio Emission}

The radio emission from classical novae was long thought to originate exclusively from thermal free-free emission generated by $\sim 10^{4}$ K photo-ionized ejecta (e.g.~\citealt{Hjellming+79,Cunningham+15}).  However, a growing sample of novae show additional maxima in their early-time radio light curves when the ejecta radius is still small and thus the brightness temperature $\gg 10^{4}$ K, which indicate an additional contribution of non-thermal synchrotron emission (e.g.~\citealt{Mioduszewski&Rupen04,Weston+16,Vlasov+16}).  

We calculate the synchrotron emission from the forward shock\footnote{Radio emission from the reverse shock is strongly attenuated by free-free absorption in the dense cool shell and is therefore negligible in comparison to the forward shock emission in a 1D model.} in our models via the following procedure.  The observed brightness temperature at radio frequency $\nu$ is given by
\begin{eqnarray}
T_{\rm B,\nu}^{\rm obs}(t) = T_{\rm B,\nu}(t)\exp[-\tau_{\rm r}(t)],
\end{eqnarray}
where $\tau_{\rm r}(t)$ is the free-free optical depth from the shock surface to infinity (see below) and $T_{\rm B,\nu}(t)$ is the un-attenuated brightness temperature of synchrotron emission from the post-shock gas.  We approximate the latter by the expression (\citealt{Vlasov+16}, their eq.~40)
\begin{eqnarray}
&& T_{\rm B,\nu} \simeq 4.2\cdot 10^{5}\,{\rm K}\,\left(\frac{n}{10^{7}\,{\rm cm^{-3}}}\right)^{\frac{1+p}{4}}\left(\frac{\nu}{10\,{\rm GHz}}\right)^{-\frac{p+3}{2}} \nonumber \\
&& \times \left(\frac{v_{\rm sh}}{10^{3}\,{\rm km\,s^{-1}}}\right)^{\frac{11+p}{2}}\left(\frac{\epsilon_{B}}{10^{-2}}\right)^{\frac{p+1}{4}}\left(\frac{\epsilon_{e}}{10^{-2}}\right)\left(\frac{\Lambda (T_{\rm sh})}{10^{-22}\,{\rm erg\, cm^3\, s^{-1}}}\right)^{-1} ,
\label{eq:TB}
\end{eqnarray}
where $\Lambda(T_{\rm sh})$ is the cooling function evaluated at the post-shock temperature $T_{\rm sh}$; $n \equiv \rho/m_p$ is the density of the gas just upstream of the shock; $v_{\rm sh}$ is the shock velocity; $\epsilon_{B}/\epsilon_{e}$ are the fraction of the shock power placed into magnetic fields and relativistic electrons, respectively, normalized to characteristic values; $p$ is the index of the energy distribution of the accelerated electrons (e.g.~$dN_e/dE \propto E^{-p}$), which we take to be $p = 2.5$.\footnote{Although synchrotron emission can also originate from secondary electron/positron pairs downstream of the shock by charged pion decay in the hadronic model for the gamma-rays, \citep{Vlasov+16} show that the observed emission is likely dominated by primary electrons directly accelerated at the shock (i.e. the radio emission is ``leptonic" even if the gamma-rays are ``hadronic").}   In deriving equation (\ref{eq:TB}) we have assumed that the shock is radiative, such that the width of the synchrotron-emitting region behind the shock responsible for the synchrotron emission is controlled by the thickness of the post-shock cooling layer $\propto 1/\Lambda$.  Since the cooling layer thickness scales as $t_{\rm cool}/t_{\rm exp}$, Figure \ref{fig:eta} shows that assuming the shocks are radiative is good for the first $\approx 40$ days in our low-mass external medium model ($r_0 = 2 \times 10^{13}$ cm) and substantially longer in the fiducial case ($r_0 = 6 \times 10^{13}$ cm).

The optical depth due to free-free absorption through the ejecta is given by  (\citealt{Rybicki&Lightman79})
\be \tau_{\rm r} = \int_{r}^{\infty}\alpha_{\rm r}dr; \,\,\,\,\alpha_{\rm r} \approx 0.06 T^{-3/2}n^{2}\nu^{-2}\,{\rm cm^{-1}},
\ee
where $T$, $n$, and $\nu$ are in cgs units and the pre-factor of $\alpha_r$ depends only weakly on the  ejecta composition \citep{Metzger+14}.  We calculate the ionization state of the ejecta ahead of the shock at each timestep following the procedure described in Appendix \ref{app:radio}.  In practice, complete ionization is a reasonable approximation at late times when $\tau_{\rm r} \lesssim 1$ given the large Lyman continuum flux from the UV radiation of the radiative shock incident on the dilute upstream medium.  Finally, we convert brightness temperature back to radio luminosity using
\be
\nu L_{\nu} = 4\pi^{2}R_{\rm sh}^{2}\frac{2\nu^{3} kT^{\rm obs}_{\rm B,\nu}}{c^{2}},
\ee
where $R_{\rm sh}$ is the shock radius.  Again, the luminosity we present assumes a spherical outflow and therefore should be multiplied by $f_{\Omega} \sim 0.3$ to account for the smaller angular extent of the equatorial shock region.  

In addition to the non-thermal synchrotron emission, we calculate the thermal free-free radiation from the hot photo-ionized gas following the procedure described in Appendix \ref{app:radio}.  Except during brief intervals when the radio photosphere is passing through a shock, the brightness temperature of the thermal emission is at the temperature floor $10^{4}$ K of the simulation.  This is reasonable physically because the latter is close to the temperature expected for gas photo-ionized by UV/X-rays from the shocks or the central white dwarf \citep{Cunningham+15}.  Also note that the light curve shape from the true bipolar ejecta morphology, including both the fast polar and slower equatorial components, will differ in detail from the predictions of a 1D model presented here (e.g.~\citealt{Ribeiro+14}).

\begin{figure}
    \centering
    \includegraphics[width=0.5\textwidth]{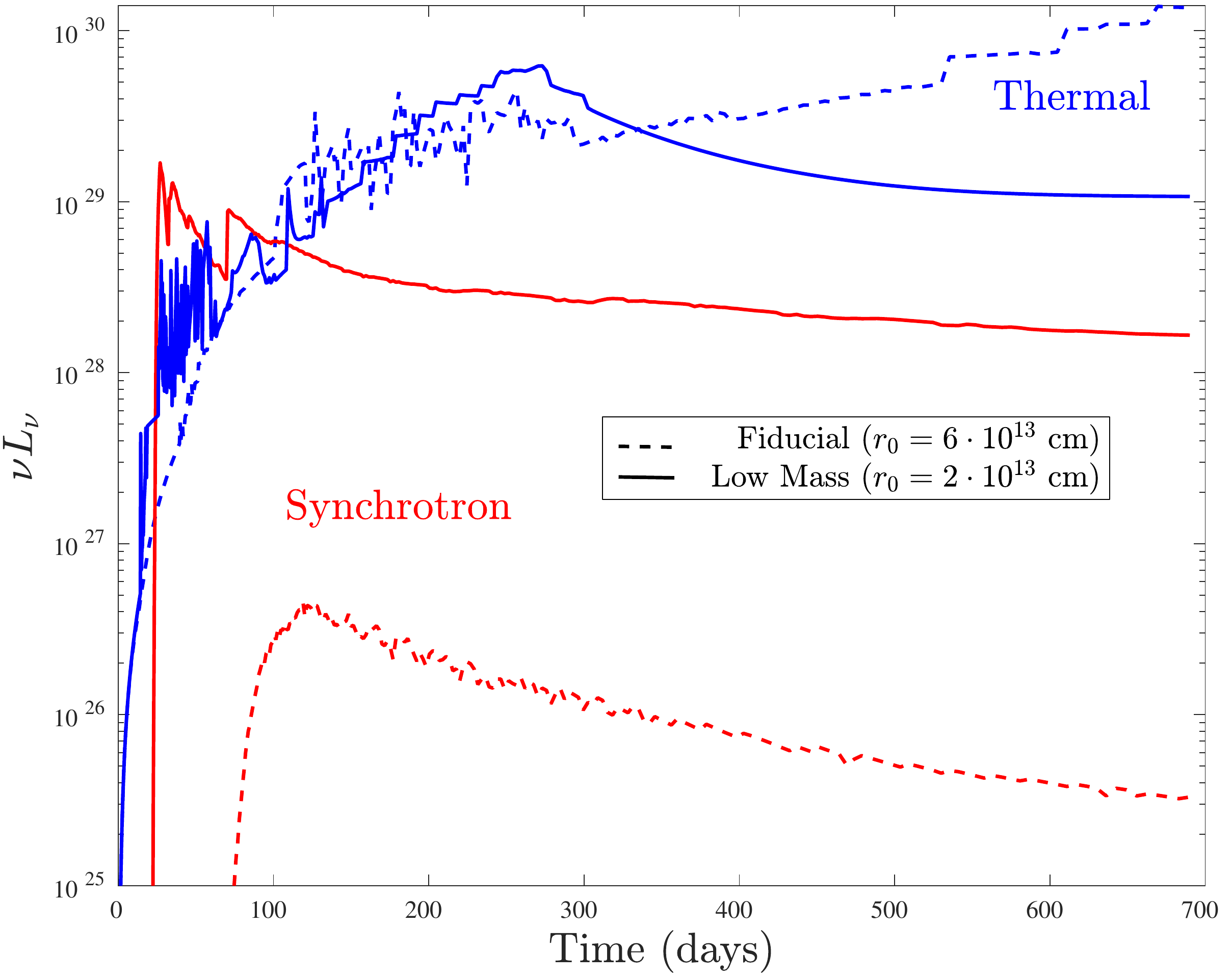}
    \includegraphics[width=0.5\textwidth]{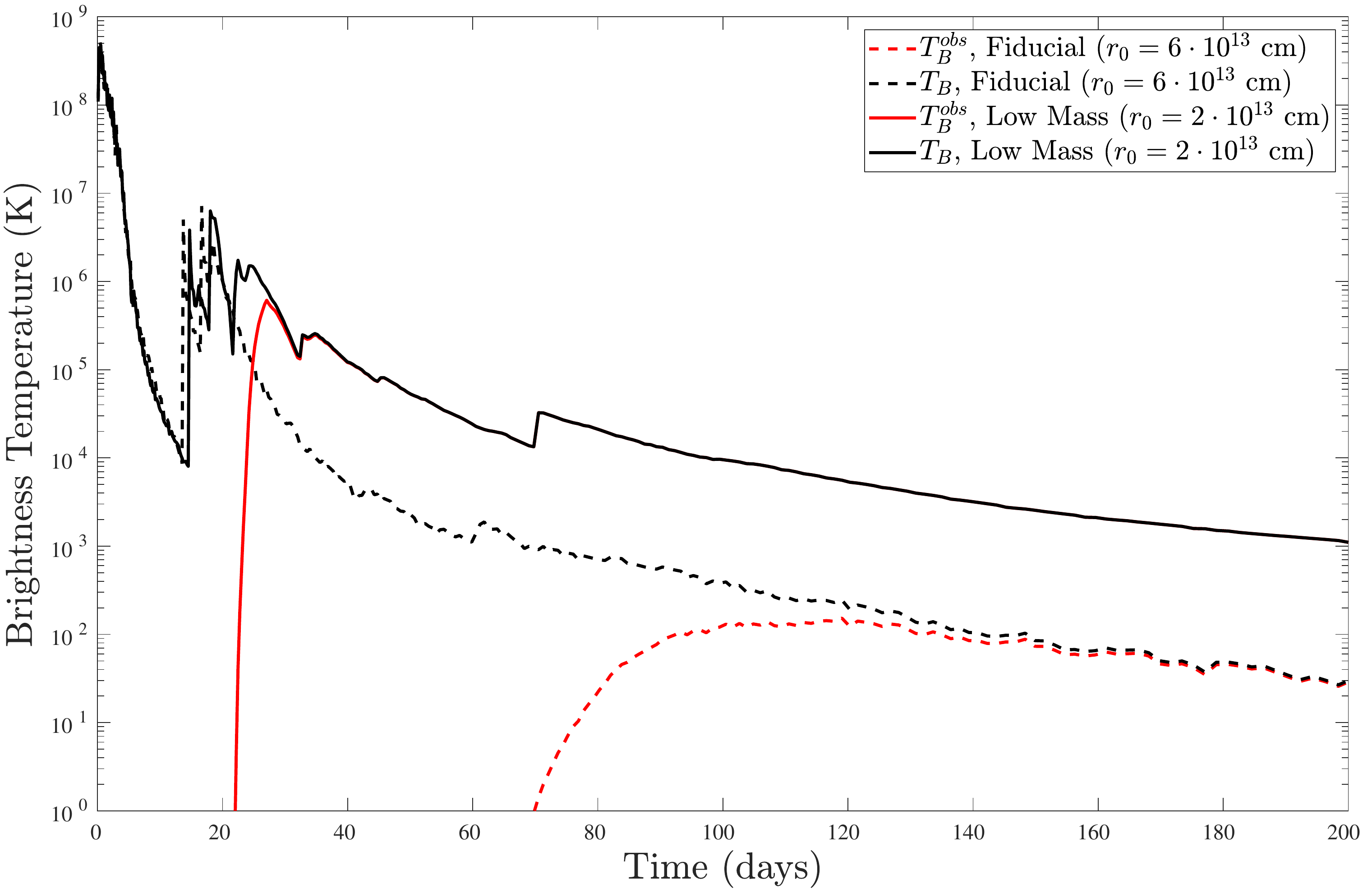}
    \caption{{\it Top:} 10 GHz radio light curves for the multiple-transition model, showing separately contributions from shock-powered synchrotron emission ({\it blue lines}) and thermal free-free emission ({\it red lines}).  We show results separately in cases in which the mass of the slow external medium is relatively high (Fiducial; $r_{0} = 6\times 10^{13}$ cm) or low (Low External Mass; $r_0 = 2\times 10^{13}$ cm).  The synchrotron emission is calculated assuming standard values for the micro-physical parameters $\epsilon_e = \epsilon_B = 10^{-2}$.  {\it Bottom:} Brightness temperatures of the forward shock synchrotron emission for the same models shown in the top panel.  We show separately the intrinsic brightness temperature at the shock, $T_{\rm B}$ ({\it black lines}), as well as the observed value, $T_{\rm B}^{\rm obs}$ ({\it red lines}), after accounting for free-free absorption by gas ahead of the shocks.}
    \label{fig:radio}
\end{figure}

\begin{figure}
    \centering
    \includegraphics[width=0.5\textwidth]{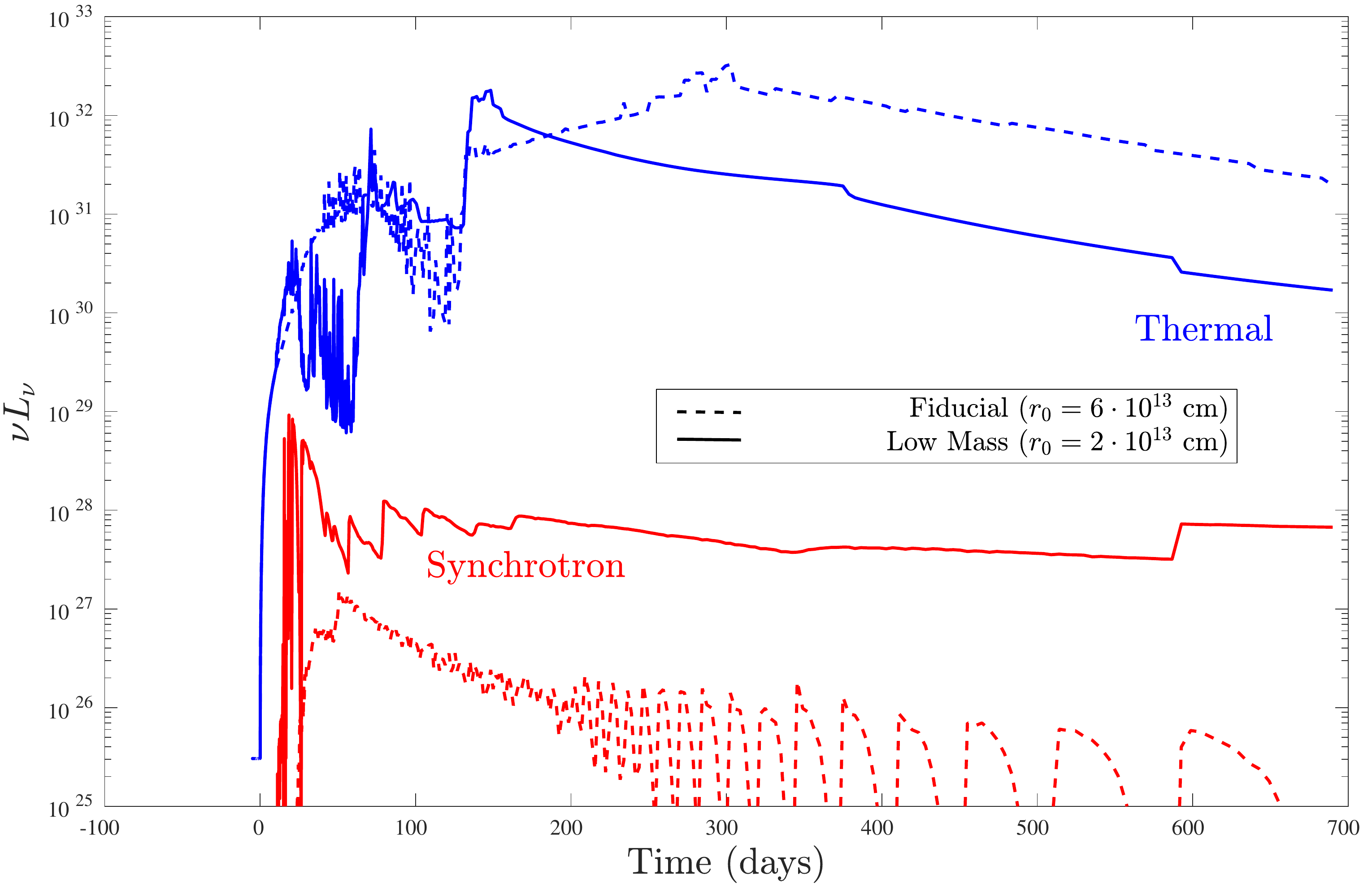}
    \caption{Same as the top panel of Fig.~\ref{fig:radio}, but for a higher radio frequency $\nu = 100$ GHz close to the ALMA bands.}
    \label{fig:radio_100}
\end{figure}

Figure \ref{fig:radio} shows the 10 GHz radio light curves (top panel) and synchrotron brightness temperatures (bottom panel) for the multiple-transition model.  We show results separately for the Fiducial ($r_0 = 6\times 10^{13}$ cm) and Low Mass ($r_0 = 2\times 10^{13}$ cm) cases.  In both models, the thermal emission ({\it blue lines}) peak after roughly one year, consistent with that of observed thermal emission in novae (e.g.~\citealt{Hjellming+79,Chomiuk+14}).  By contrast, the non-thermal synchrotron emission ({\it red lines}) varies by two orders of magnitude between the models.  The synchrotron component furthermore only peaks above the thermal free-free emission in the Low External Mass model ($r_0 = 2\times 10^{13}$ cm).  Figure \ref{fig:radio_100} shows similar light curves, but in the case of a higher frequency $\nu = 100$ GHz.  In this case the synchrotron emission is overwhelmed by the thermal component at all times in both models.  

As in the X-ray case, although the power of the shocks peaks at early times, the peak in the synchrotron light curve is delayed by free-free absorption through the slow external medium until $\tau_{\rm r} \lesssim 1$.  The wide range in the peak luminosity of the shocks results because of the sensitive dependence of the intrinsic synchrotron luminosity on the shock velocity ($T_{\rm B} \propto v_{\rm sh}^{6.8}$ for $ p \approx 2.5$; eq.~\ref{eq:TB}).  The value of $v_{\rm sh}$ at late times is in turn sensitive to the amount of slow ejecta into which the outermost shock is decelerating.  The large diversity in the predicted peak luminosities of the non-thermal radio emission appears broadly compatible with the range $\nu L_{\nu} \lesssim 3\times 10^{28}$ erg s$^{-1}$ \citep{Chomiuk+14} to  $\sim 3\times 10^{30}$ erg s$^{-1}$ \citep{Weston+16} from observations given realistic variations in the properties of the external medium and of the fast outflow from the white dwarf.

\subsection{Dust Formation}
\label{sec:dust}

Dust formation is commonly inferred to take place in the ejecta of novae (e.g.~\citealt{Gehrz+98}).  This is evidenced by the delayed onset of infrared emission and the occurrence in many cases of deep minima in the optical light curves caused by the sudden formation of dust along the line of site (e.g.~\citealt{Evans+05,Strope+10,Evans+17,Harvey+18,Finzell+18,Aydi+19}). 
The expanding ejecta becomes cool enough for solid condensation to take place within weeks to months of the outburst, once the ejecta reaches radii larger than,
\begin{equation}
    r_{\rm dust} =\left(\frac{L}{4\pi\sigma T_{\rm cond}^4}\right)^{1/2},
\label{eq:rdust}
\end{equation}
where $T=T_{\rm cond} \approx 1500$ K is the approximate condensation temperature and $L$ is the central irradiating luminosity.  Even once the condition $r > r_{\rm dust}$ is satisfied, however, it is not clear how molecules and grains are able to form given the potentially harsh environment created by ionizing radiation from the white dwarf (e.g.~\citealt{Pontefract&Rawlings04}), particularly if the ejecta shell is part of a smooth temporally-extended (and thus radially thick) outflow.  

\begin{figure}
    \centering
    \includegraphics[width=0.5\textwidth]{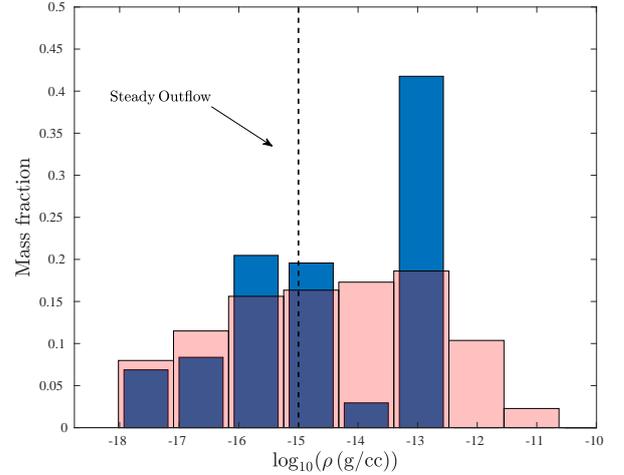}
    \caption{Radiative shocks in novae provide a natural mechanism to produce clumpy ejecta structure with a spread in densities far in excess of that predicted in a smoothly-evolving outflow.  Here we show the mass-weighted distribution of the gas density of radiative shocks in our single-transition ({\it red }) and multiple-transition ({\it blue}) models as measured at the time of solid condensation ($r = r_{\rm dust}$, eq.~\ref{eq:rdust}), calculated assuming a central white dwarf luminosity $L = 10^{38}$ erg s$^{-1}$.  The distributions shown are obtained by convolving the density distribution measured from our 1D radiative shock models with a lognormal density distribution motivated by the results of zoomed-in high resolution simulations of radiative shocks from \citet{Steinberg&Metzger18}, which capture the clumpy structure generated by thermal instabilities.  Shown for comparison with a vertical dashed line is the average density for an otherwise similar outflow of a constant mass-loss rate $\dot{M}$ without shocks (eq.~\ref{eq:rhowind}).}
    \label{fig:dust}
\end{figure}

\citet{Derdzinski+17} argue that the cool, dense gas generated by radiative shocks provide ideal conditions capable of dust nucleation and growth (see also \citealt{Aydi+19}).  We expand upon this possibility explicitly here in Figure \ref{fig:dust}, which shows the mass-weighted distribution of gas density for matter passing through the condensation radius $r_{\rm dust}$ in our single and multiple-transition models.  Our 1D simulations cannot capture the full multi-dimensional, multi-phase structure of the thermally-unstable post-shock gas.  To generate Figure \ref{fig:dust} we have therefore convolved the relatively narrow density distribution derived from our 1D models with a lognormal distribution (with a mean $\sim$ variance $\sim \mathcal{M}^{2}$) motivated by the results of high-resolution 2D simulations of high Mach number $\mathcal{M} \gg 1$ radiative shocks in \citet{Steinberg&Metzger18}.  These simulations capture the wide range of gas densities in the multi-phase region generated by thermal instabilities in the post-shock gas.

It is fruitful to compare the density distribution of shocked gas to the alternative simpler scenario of a steady outflow in which no shocks occur.  For an outflow of constant mass-loss rate $\dot{M}$, the outflow density at $r = r_{\rm dust}$ is
\begin{eqnarray}
    \rho_{\rm w} &=& \frac{\dot{M}\sigma T_{\rm cond}^4}{L v_w} \nonumber \\
&\approx& 10^{-15}\,{\rm g\,cm^{-3}}\left(\frac{\dot{M}}{10^{-5}M_\odot/\textrm{wk}}\right)\left(\frac{L}{10^{38}\,\rm erg/s}\right)^{-1}\left(\frac{v_w}{10^{3}\,\rm km/s}\right)^{-1}. \label{eq:rhowind}
\end{eqnarray}
Comparing a characteristic value of $\rho_{\rm w}$ (shown as a vertical dashed line in Fig.~\ref{fig:dust}) to that obtained in the case of radiative shocks for a flow with otherwise similar mean mass outflow rates, we see that the shock case spans $\sim 4-5$ orders of magnitude in density.  

The ``clumpy'' ejecta structure generated in the thin shells of radiative internal shocks could provide a natural explanation for the small average filling factors inferred in nova ejecta of $\lesssim$ few percent (e.g.~\citealt{Shore+13}), which otherwise would be challenging to produce in a steady-wind or the smooth homologously-expanding outflow from a singular mass ejection event.  These high-density pockets also provide local environments for the nucleation and growth of dust grains as they are shielded from the otherwise detrimental affects of X-ray/UV radiation \citep{Derdzinski+17}.  

\section{Conclusions}
\label{sec:conclusions}

\begin{figure}
    \centering
 \includegraphics[width=0.5\textwidth]{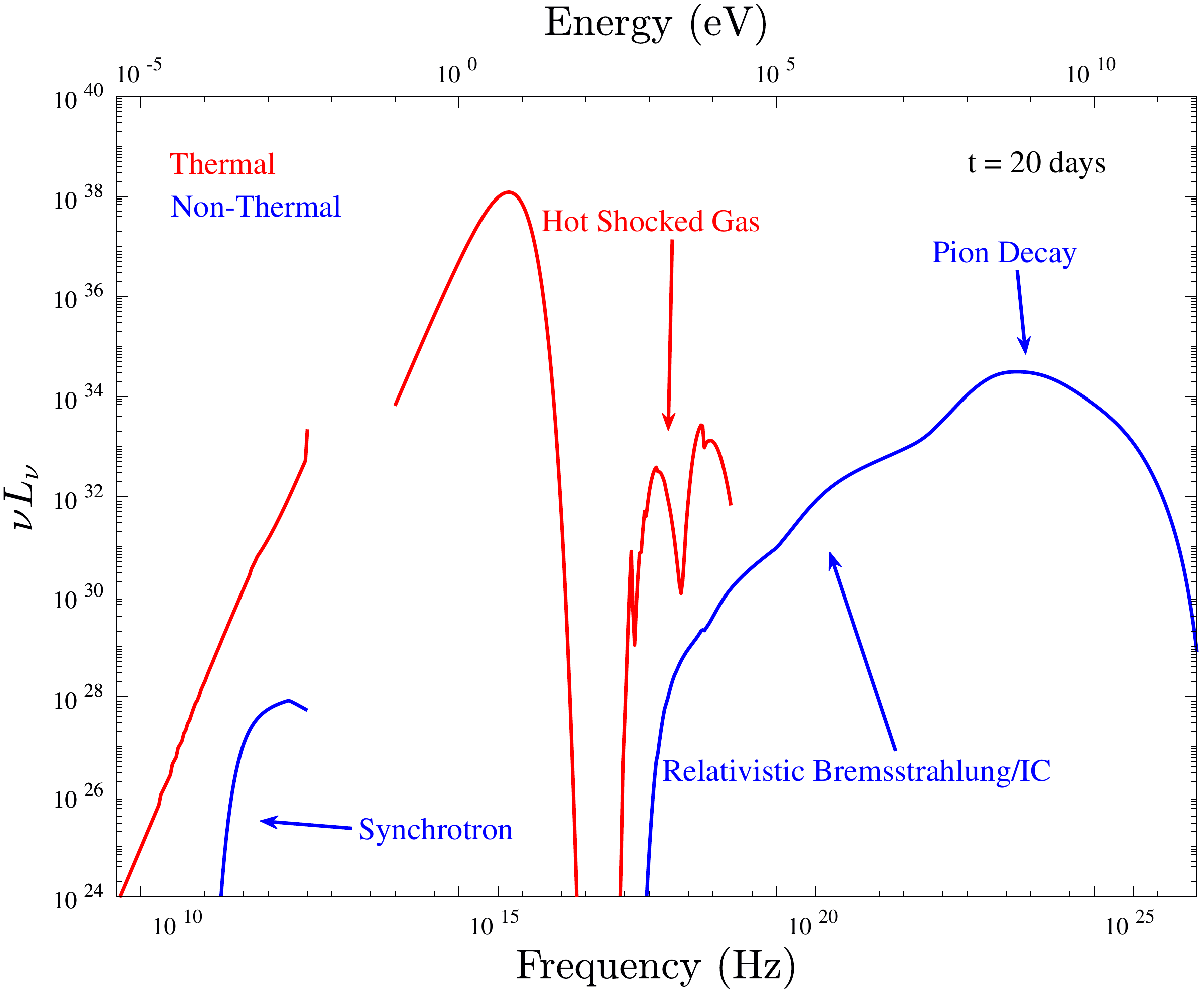}
    \includegraphics[width=0.5\textwidth]{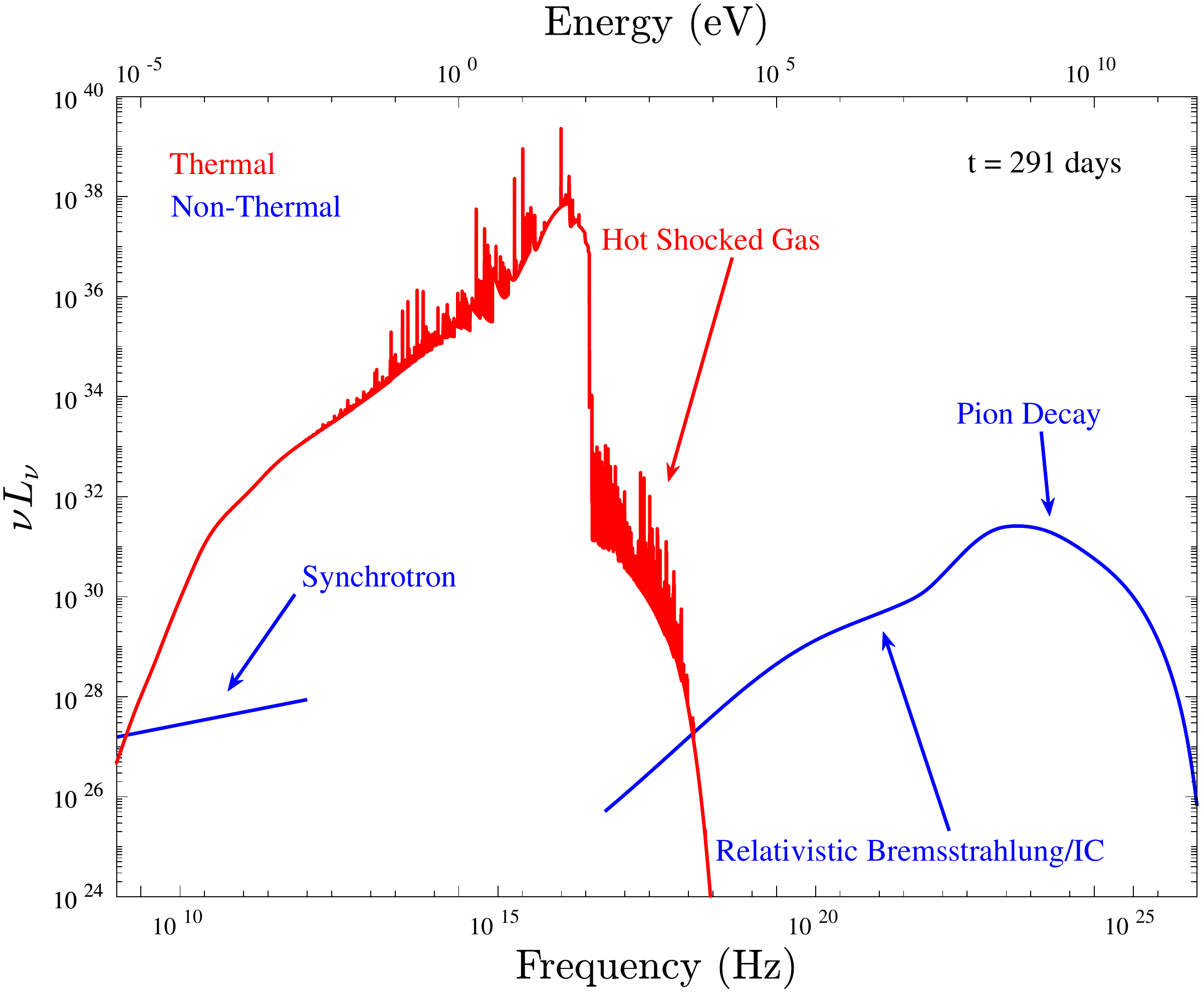}
    \caption{Schematic spectra energy distribution, $\nu L_{\nu}$,  at two snapshots $t = 20$ days ({\it Top}) and $t = 291$ days ({\it Bottom}) in the Multiple-Transition Low External Mass model.  Red lines show thermal emission components, including (1) X-ray emission which escapes directly from the shocks despite photo-electric absorption by the overlying ejecta; (2) emission from the white dwarf and internal shocks which has been absorbed and reprocessed into the optical/UV band, emerging either as optically-thick emission at early times ($t = 20$ day) or as an optically-thin continuum/emission lines at late times ($t = 291$ days).  We calculate the latter by post-processing our ejecta profiles through {\it CLOUDY}.  Solar abundances are assumed in calculating the gas cooling and ejecta opacity.  Blue lines show non-thermal emission components, including (1) radio synchrotron emission from the outermost forward shock; (2) gamma-rays from pion decay (and relativistic bremsstrahlung from secondary pairs) generated by relativistic ions accelerated at the shocks (the spectrum shown was taken from the best-fit model to the LAT emission from ASASSN15-lh from \citealt{Li+17} following \citealt{Vurm&Metzger18}).}
    \label{fig:sed}
\end{figure}

We have studied internal shocks in classical nova outflows by means of 1D hydrodynamical simulations including radiative losses.  Our primary aim is to model the shock interaction between slow and fast outflows from these systems and to assess the ability of shocks to simultaneously explain multi-wavelength observations.  Figure \ref{fig:sed} summarizes the broad range of emission components (both thermal and non-thermal) which contribute to the spectral energy distribution of novae, many of which are fundamentally shock-powered. 

Our conclusions may be summarized as follows:
\begin{itemize}
\item{The hourglass/embedded ring geometry of nova ejecta on large spatial scales is suggestive of earlier shock interaction between a fast outflow from the white dwarf with a slower equatorially-concentrated outflow shaped by the binary orbit (Fig.~\ref{fig:cartoon}).  However, this structure would be relatively insensitive to the detailed time evolution of the outflow at early times, in particular whether the transition from slow-to-fast outflow modes occurs only once (``single-transition'' scenario) or several times in succession (``multiple-transition'' scenario). }
\item{The interaction between consecutive fast and slow outflow components is mediated by a forward-reverse shock structure, which sweeps up and compresses the shocked gas into a narrow dense shell.  Each transition back to a fast flow generates its own dense shell.  In most cases these shells merge pairwise into a single monolithic shell, which continues to propagate down the density gradient of the initial slow outflow until late times (Fig.~\ref{fig:snapshot}).}
\item{For the first $\sim$ month or longer after the outburst, the shocks take place deep under the UV and soft X-ray photosphere.  Direct radiation from the shock is absorbed and reprocessed into optical emission, powering time-variable optical luminosity on top of the more smoothly-varying light curve powered by nuclear burning on the white dwarf surface (Fig.~\ref{fig:Ltot}).  In single-transition scenarios, the shock quickly moves ahead of the optical photosphere, such that after days or less its reprocessed luminosity is channeled into emission lines or optically-thin emission.     

By contrast, the fresh shocks continually generated at small radii in the multiple transition scenario result in much of the shock emission taking place below the photosphere, where it can contribute to the optically-thick continuum.  This behavior leads to correlated maxima in the optical light curve and photospheric radii, with a possible short time lag of days or less, consistent with observations of some flares in novae (e.g.~\citealt{Tanaka+11a,Aydi+19}).  For shock-powered optical emission above the photosphere, the predicted evolution of the spectral line velocities can be complex, with accelerating, decelerating, and even merging velocity components (Fig.~\ref{fig:dL_dv}).  The emission lines originate from material behind the forward and reverse shocks with a velocity width intermediate between those of the slowest and fastest outflows.  The slow unshocked medium may instead be that which is probed by transient narrow absorption line features (e.g. the THEA systems described by \citealt{Williams+08}).
}
\item{Thermal instabilities behind radiative shocks naturally produce a clumpy medium in the post-shock central shells, spanning several orders of magnitude in the gas density (Fig.~\ref{fig:dust}).  \citet{Mason+18} argue, from high-resolution UV and optical spectroscopy, that gamma-ray producing novae can be understood  by a clumpy ejecta shell expelled during a brief, singular ejection episode, rather than necessitating a wind-type geometry.  Our models predict that the majority (of at least the equatorial component) of the ejecta has passed through at least one radiative shock on its journey to large radii.  Such clumpiness is therefore a natural prediction of the model.  The same clumpy thin shells provide natural sites for dust nucleation, once the ejecta shells reach sufficiently large radii for solid condensation.  }
\item{As both gamma-ray and optical emission track the shock power with little or no delay, correlated optical and gamma-ray flares are a natural prediction of the Multiple Transition model in which a significant fraction of the optical luminosity originates from shocks.  The observed range of luminosities are obtained if the acceleration efficiency of relativistic particles at the shock is $\sim 0.3\%$, similar to those expected due to ion acceleration at the corrugated interface of radiative shocks \citep{Steinberg&Metzger18} and thus further supporting a hadronic origin for the gamma-ray emission (e.g.~\citealt{Metzger+16,Li+17,Martin+18}).}  
\item{The Multiple Transition scenario is clearly favored by the gamma-ray/optical variability of novae.  However, this does not address the physical origin of the abrupt and frequent transitions in the nova outflow properties between ``fast" and ``slow" modes, over a wide range of timescales.  The fast outflow likely originates directly from the white dwarf, while$-$given its low velocity and equatorially-focused geometry$-$the slow outflow probably originates from Roche lobe overflow (by frictional heating, non-axisymmetric gravitational torques from the binary, or in a wind from a circumbinary excretion disk; \citealt{Pejcha+16a,Pejcha+16b}).  \citet{Kato&Hachisu11} model the structure of the white dwarf burning layer accounting in an approximate matter for the effects of the binary companion.  They find distinct optically-thick wind (``fast") and static envelope (``slow") phases, transitions between which could give rise to the observed variability (and, in our picture, internal shocks).  However, to advance to the stage of being truly predictive (e.g. regarding the timescale between outflow mode switches), models of novae must account self-consistently for the multi-dimensional boundary conditions imposed by the Roche Lobe and identify the dominant physical mechanism driving the slow outflow.}  .

\item{Due to the relatively low photoelectric opacity at higher photon energies, flares in the hard $\gtrsim 10$ keV band may accompany the optical/gamma-ray flares (Figs.~\ref{fig:xray}, \ref{fig:xray2}).  This is particularly true in cases in which external medium has a low metallicity and solar-like composition (e.g. ejecta dominated by unburnt material from the companion star).  On the other hand, the hard X-ray emission could be substantially suppressed from the naive predictions of the 1D jump conditions if the shocks remain radiative to late times \citep{Steinberg&Metzger18}.}
\item{In contrast to the hard X-rays, the soft $\lesssim 1$ keV emission is typically delayed for weeks or longer by photoelectric absorption in the external slow outflow.  This emission originates not from the deeply-embedded shell collisions themselves, but from the forward shock of the outermost shell, once it reaches a sufficiently low external column.  The luminosity range of the $\sim $keV X-ray emission spans several orders of magnitude, being particularly sensitive to the  mass, composition, and radial distribution of the initial slow outflow, or pre-existing medium surrounding the binary, at the time of the first transition to a fast outflow.  In some cases the X-ray light curve may peak at sufficiently low energies $\lesssim 0.3$ keV that it could be confused for emission from the white dwarf surface in the absence of good spectral coverage.} 
\item{Like the soft X-rays, radio synchrotron emission from shock-accelerated electrons is not generally visible until late times $\gtrsim$ weeks-months (in this case due to free-free absorption) from the forward shock of the outermost shell (Fig.~\ref{fig:radio}).  If the forward shock is radiative, the radio emission is extremely sensitive to the velocity of the forward shock, which in turn depends on the momentum input of the fast outflow and the mass profile of the initial slow outflow or pre-existing external medium.  

The keV X-ray and synchrotron radio emission together provide a potentially sensitive probe of the initial  (potentially pre-thermonuclear runaway) medium surrounding the white dwarf, the origin of which is debated (e.g.~\citealt{Williams+08}).  Unfortunately detailed inferences may be plagued by several degeneries, such as the microphysical parameters $\epsilon_e$, $\epsilon_B$ in the radio case and the precise ejecta composition in the X-ray case.  Very early time radio detections, especially at low frequencies where free-free absorption is severe, cannot be explained by the slow equatorial outflow and may require an additional emission site, such internal shocks in the polar regions due to variability in the fast outflow itself (\citealt{Vlasov+16}).}
\end{itemize}

\section*{Acknowledgements}
We thank Elias Aydi, Laura Chomiuk, and Indrek Vurm for helpful information and discussions.  This work is supported in part by NSF through grant \#AST-1615084 and by NASA through the {\it Fermi} Guest Investigator Program (grant \# 80NSSC18K1708).  BDM received partial support from the Simons Foundation through the Simons Fellows Program (grant \# 606260).

\bibliography{novashocks}

\begin{thebibliography}{87}
\providecommand\natexlab[1]{#1}
\providecommand\JournalTitle[1]{#1}

\bibitem[{{Ackermann} {et~al.}(2014)}]{Ackermann+14}
{Ackermann}, M., {et~al.} 2014,
  \href{http://dx.doi.org/10.1126/science.1253947}{\JournalTitle{Science}, 345,
  554}

\bibitem[{{Aydi} {et~al.}(2019){Aydi}, {Chomiuk}, {Strader}, {Swihart},
  {Bahramian}, {Harvey}, {Britt}, {Buckley}, {Chen}, {Dage}, {Darnley}, {Dong},
  {Hambsch}, {Holoien}, {Jha}, {Kochanek}, {Kuin}, {Li}, {Monard}, {Mukai},
  {Page}, {Prieto}, {Richardson}, {Shappee}, {Shishkovsky}, {Sokolovsky},
  {Stanek}, \& {Thompson}}]{Aydi+19}
{Aydi}, E., {Chomiuk}, L., {Strader}, J., {et~al.} 2019, \JournalTitle{arXiv
  e-prints}, \href{http://arxiv.org/abs/1903.09232}{{\sffamily arXiv:1903.09232
  [astro-ph.HE]}}

\bibitem[{{Bode} \& {Evans}(2008)}]{Bode&Evans08}
{Bode}, M.~F., \& {Evans}, A. 2008, {Classical Novae}

\bibitem[{{Caprioli} \& {Spitkovsky}(2014)}]{Caprioli&Spitkovsky14}
{Caprioli}, D., \& {Spitkovsky}, A. 2014,
  \href{http://dx.doi.org/10.1088/0004-637X/783/2/91}{\JournalTitle{\apj}, 783,
  91}

\bibitem[{{Casanova} {et~al.}(2018){Casanova}, {Jos{\'e}}, \&
  {Shore}}]{Casanova+18}
{Casanova}, J., {Jos{\'e}}, J., \& {Shore}, S.~N. 2018,
  \href{http://dx.doi.org/10.1051/0004-6361/201833422}{\JournalTitle{\aap},
  619, A121}

\bibitem[{{Cassatella} {et~al.}(2004){Cassatella}, {Lamers}, {Rossi},
  {Altamore}, \& {Gonz{\'a}lez-Riestra}}]{Cassatella+04}
{Cassatella}, A., {Lamers}, H.~J.~G.~L.~M., {Rossi}, C., {Altamore}, A., \&
  {Gonz{\'a}lez-Riestra}, R. 2004,
  \href{http://dx.doi.org/10.1051/0004-6361:20034102}{\JournalTitle{\aap}, 420,
  571}

\bibitem[{{Cheung} {et~al.}(2016){Cheung}, {Jean}, {Shore}, {Stawarz},
  {Corbet}, {Kn{\"o}dlseder}, {Starrfield}, {Wood}, {Desiante}, {Longo},
  {Pivato}, \& {Wood}}]{Cheung+16}
{Cheung}, C.~C., {Jean}, P., {Shore}, S.~N., {et~al.} 2016,
  \href{http://dx.doi.org/10.3847/0004-637X/826/2/142}{\JournalTitle{\apj},
  826, 142}

\bibitem[{{Chevalier} \& {Imamura}(1982)}]{Chevalier&Imamura82}
{Chevalier}, R.~A., \& {Imamura}, J.~N. 1982,
  \href{http://dx.doi.org/10.1086/160364}{\JournalTitle{\apj}, 261, 543}

\bibitem[{{Chomiuk} {et~al.}(2014){Chomiuk}, {Linford}, {Yang}, {O'Brien},
  {Paragi}, {Mioduszewski}, {Beswick}, {Cheung}, {Mukai}, {Nelson}, {Ribeiro},
  {Rupen}, {Sokoloski}, {Weston}, {Zheng}, {Bode}, {Eyres}, {Roy}, \&
  {Taylor}}]{Chomiuk+14}
{Chomiuk}, L., {Linford}, J.~D., {Yang}, J., {et~al.} 2014,
  \href{http://dx.doi.org/10.1038/nature13773}{\JournalTitle{\nat}, 514, 339}

\bibitem[{{Cs{\'a}k} {et~al.}(2005){Cs{\'a}k}, {Kiss}, {Retter}, {Jacob}, \&
  {Kaspi}}]{Csak+05}
{Cs{\'a}k}, B., {Kiss}, L.~L., {Retter}, A., {Jacob}, A., \& {Kaspi}, S. 2005,
  \href{http://dx.doi.org/10.1051/0004-6361:20035751}{\JournalTitle{\aap}, 429,
  599}

\bibitem[{{Cunningham} {et~al.}(2015){Cunningham}, {Wolf}, \&
  {Bildsten}}]{Cunningham+15}
{Cunningham}, T., {Wolf}, W.~M., \& {Bildsten}, L. 2015,
  \href{http://dx.doi.org/10.1088/0004-637X/803/2/76}{\JournalTitle{\apj}, 803,
  76}

\bibitem[{{Derdzinski} {et~al.}(2017){Derdzinski}, {Metzger}, \&
  {Lazzati}}]{Derdzinski+17}
{Derdzinski}, A.~M., {Metzger}, B.~D., \& {Lazzati}, D. 2017,
  \href{http://dx.doi.org/10.1093/mnras/stx829}{\JournalTitle{\mnras}, 469,
  1314}

\bibitem[{{Evans} {et~al.}(2005){Evans}, {Tyne}, {Smith}, {Geballe},
  {Rawlings}, \& {Eyres}}]{Evans+05}
{Evans}, A., {Tyne}, V.~H., {Smith}, O., {et~al.} 2005,
  \href{http://dx.doi.org/10.1111/j.1365-2966.2005.09146.x}{\JournalTitle{\mnras},
  360, 1483}

\bibitem[{{Evans} {et~al.}(2017){Evans}, {Banerjee}, {Gehrz}, {Joshi}, {Ashok},
  {Ribeiro}, {Darnley}, {Woodward}, {Sand}, \& {Marion}}]{Evans+17}
{Evans}, A., {Banerjee}, D.~P.~K., {Gehrz}, R.~D., {et~al.} 2017,
  \href{http://dx.doi.org/10.1093/mnras/stw3334}{\JournalTitle{\mnras}, 466,
  4221}

\bibitem[{{Ferland} {et~al.}(1998){Ferland}, {Korista}, {Verner}, {Ferguson},
  {Kingdon}, \& {Verner}}]{Ferland+98}
{Ferland}, G.~J., {Korista}, K.~T., {Verner}, D.~A., {et~al.} 1998,
  \href{http://dx.doi.org/10.1086/316190}{\JournalTitle{\pasp}, 110, 761}

\bibitem[{{Ferland} {et~al.}(2017){Ferland}, {Chatzikos}, {Guzm{\'a}n},
  {Lykins}, {van Hoof}, {Williams}, {Abel}, {Badnell}, {Keenan}, {Porter}, \&
  {Stancil}}]{Ferland+17}
{Ferland}, G.~J., {Chatzikos}, M., {Guzm{\'a}n}, F., {et~al.} 2017,
  \JournalTitle{Revista Mexicana de Astronomía y Astrofísica}, 53, 385

\bibitem[{{Finzell} {et~al.}(2018){Finzell}, {Chomiuk}, {Metzger}, {Walter},
  {Linford}, {Mukai}, {Nelson}, {Weston}, {Zheng}, {Sokoloski}, {Mioduszewski},
  {Rupen}, {Dong}, {Starrfield}, {Cheung}, {Woodward}, {Taylor}, {Bohlsen},
  {Buil}, {Prieto}, {Wagner}, {Bensby}, {Bond}, {Sumi}, {Bennett}, {Abe},
  {Koshimoto}, {Suzuki}, {Tristram}, {Christie}, {Natusch}, {McCormick}, {Yee},
  \& {Gould}}]{Finzell+18}
{Finzell}, T., {Chomiuk}, L., {Metzger}, B.~D., {et~al.} 2018,
  \href{http://dx.doi.org/10.3847/1538-4357/aaa12a}{\JournalTitle{\apj}, 852,
  108}

\bibitem[{{Franckowiak} {et~al.}(2018){Franckowiak}, {Jean}, {Wood}, {Cheung},
  \& {Buson}}]{Franckowiak+18}
{Franckowiak}, A., {Jean}, P., {Wood}, M., {Cheung}, C.~C., \& {Buson}, S.
  2018,
  \href{http://dx.doi.org/10.1051/0004-6361/201731516}{\JournalTitle{\aap},
  609, A120}

\bibitem[{{Gallagher} \& {Starrfield}(1978)}]{Gallagher&Starrfield78}
{Gallagher}, J.~S., \& {Starrfield}, S. 1978,
  \href{http://dx.doi.org/10.1146/annurev.aa.16.090178.001131}{\JournalTitle{\araa},
  16, 171}

\bibitem[{{Gehrz} {et~al.}(1998){Gehrz}, {Truran}, {Williams}, \&
  {Starrfield}}]{Gehrz+98}
{Gehrz}, R.~D., {Truran}, J.~W., {Williams}, R.~E., \& {Starrfield}, S. 1998,
  \href{http://dx.doi.org/10.1086/316107}{\JournalTitle{\pasp}, 110, 3}

\bibitem[{{Gill} \& {O'Brien}(2000)}]{Gill&OBrien00}
{Gill}, C.~D., \& {O'Brien}, T.~J. 2000,
  \href{http://dx.doi.org/10.1046/j.1365-8711.2000.03342.x}{\JournalTitle{\mnras},
  314, 175}

\bibitem[{{Gnat} \& {Sternberg}(2007)}]{Gnat07}
{Gnat}, O., \& {Sternberg}, A. 2007,
  \href{http://dx.doi.org/10.1086/509786}{\JournalTitle{\apjs}, 168, 213}

\bibitem[{{Harman} \& {O'Brien}(2003)}]{Harman&OBrien03}
{Harman}, D.~J., \& {O'Brien}, T.~J. 2003,
  \href{http://dx.doi.org/10.1046/j.1365-8711.2003.06906.x}{\JournalTitle{\mnras},
  344, 1219}

\bibitem[{{Harvey} {et~al.}(2018){Harvey}, {Redman}, {Darnley}, {Williams},
  {Berdyugin}, {Piirola}, {Fitzgerald}, \& {O'Connor}}]{Harvey+18}
{Harvey}, E.~J., {Redman}, M.~P., {Darnley}, M.~J., {et~al.} 2018,
  \href{http://dx.doi.org/10.1051/0004-6361/201731741}{\JournalTitle{\aap},
  611, A3}

\bibitem[{{Henze} {et~al.}(2014){Henze}, {Ness}, {Darnley}, {Bode}, {Williams},
  {Shafter}, {Kato}, \& {Hachisu}}]{Henze+14}
{Henze}, M., {Ness}, J.-U., {Darnley}, M.~J., {et~al.} 2014,
  \href{http://dx.doi.org/10.1051/0004-6361/201423410}{\JournalTitle{\aap},
  563, L8}

\bibitem[{{Henze} {et~al.}(2018){Henze}, {Darnley}, {Williams}, {Kato},
  {Hachisu}, {Anupama}, {Arai}, {Boyd}, {Burke}, {Ciardullo}, {Chinetti},
  {Cook}, {Cook}, {Erdman}, {Gao}, {Harris}, {Hartmann}, {Hornoch}, {Horst},
  {Hounsell}, {Husar}, {Itagaki}, {Kabashima}, {Kafka}, {Kaur}, {Kiyota},
  {Kojiguchi}, {Ku{\v c}{\'a}kov{\'a}}, {Kuramoto}, {Maehara}, {Mantero},
  {Masci}, {Matsumoto}, {Naito}, {Ness}, {Nishiyama}, {Oksanen}, {Osborne},
  {Page}, {Paunzen}, {Pavana}, {Pickard}, {Prieto-Arranz},
  {Rodr{\'{\i}}guez-Gil}, {Sala}, {Sano}, {Shafter}, {Sugiura}, {Tan},
  {Tordai}, {Vra{\v s}til}, {Wagner}, {Watanabe}, {Williams}, {Bode}, {Bruno},
  {Buchheim}, {Crawford}, {Goff}, {Hernanz}, {Igarashi}, {Jos{\'e}}, {Motta},
  {O'Brien}, {Oswalt}, {Poyner}, {Ribeiro}, {Sabo}, {Shara}, {Shears},
  {Starkey}, {Starrfield}, \& {Woodward}}]{Henze+18}
{Henze}, M., {Darnley}, M.~J., {Williams}, S.~C., {et~al.} 2018,
  \href{http://dx.doi.org/10.3847/1538-4357/aab6a6}{\JournalTitle{\apj}, 857,
  68}

\bibitem[{{Hillman} {et~al.}(2014){Hillman}, {Prialnik}, {Kovetz}, {Shara}, \&
  {Neill}}]{Hillman+14}
{Hillman}, Y., {Prialnik}, D., {Kovetz}, A., {Shara}, M.~M., \& {Neill}, J.~D.
  2014, \href{http://dx.doi.org/10.1093/mnras/stt2027}{\JournalTitle{\mnras},
  437, 1962}

\bibitem[{{Hjellming} {et~al.}(1979){Hjellming}, {Wade}, {Vandenberg}, \&
  {Newell}}]{Hjellming+79}
{Hjellming}, R.~M., {Wade}, C.~M., {Vandenberg}, N.~R., \& {Newell}, R.~T.
  1979, \href{http://dx.doi.org/10.1086/112585}{\JournalTitle{\aj}, 84, 1619}

\bibitem[{{Jack} {et~al.}(2017){Jack}, {Robles P{\'e}rez}, {De Gennaro Aquino},
  {Schr{\"o}der}, {Wolter}, {Eenens}, {Schmitt}, {Mittag}, {Hempelmann},
  {Gonz{\'a}lez-P{\'e}rez}, {Rauw}, \& {Hauschildt}}]{Jack+17}
{Jack}, D., {Robles P{\'e}rez}, J.~d.~J., {De Gennaro Aquino}, I., {et~al.}
  2017,
  \href{http://dx.doi.org/10.1002/asna.201613217}{\JournalTitle{Astronomische
  Nachrichten}, 338, 91}

\bibitem[{{Kato} \& {Hachisu}(1991)}]{Kato&Hachisu91}
{Kato}, M., \& {Hachisu}, I. 1991,
  \href{http://dx.doi.org/10.1086/170081}{\JournalTitle{\apj}, 373, 620}

\bibitem[{{Kato} \& {Hachisu}(1994)}]{Kato&Hachisu94}
---. 1994, \href{http://dx.doi.org/10.1086/175041}{\JournalTitle{\apj}, 437,
  802}

\bibitem[{{Kato} \& {Hachisu}(2009)}]{Kato&Hachisu09}
---. 2009,
  \href{http://dx.doi.org/10.1088/0004-637X/699/2/1293}{\JournalTitle{\apj},
  699, 1293}

\bibitem[{{Kato} \& {Hachisu}(2011)}]{Kato&Hachisu11}
---. 2011,
  \href{http://dx.doi.org/10.1088/0004-637X/743/2/157}{\JournalTitle{\apj},
  743, 157}

\bibitem[{{Kee} {et~al.}(2014){Kee}, {Owocki}, \& {ud-Doula}}]{Kee+14}
{Kee}, N.~D., {Owocki}, S., \& {ud-Doula}, A. 2014,
  \href{http://dx.doi.org/10.1093/mnras/stt2475}{\JournalTitle{\mnras}, 438,
  3557}

\bibitem[{{Krauss} {et~al.}(2011){Krauss}, {Chomiuk}, {Rupen}, {Roy},
  {Mioduszewski}, {Sokoloski}, {Nelson}, {Mukai}, {Bode}, {Eyres}, \&
  {O'Brien}}]{Krauss+11}
{Krauss}, M.~I., {Chomiuk}, L., {Rupen}, M., {et~al.} 2011,
  \href{http://dx.doi.org/10.1088/2041-8205/739/1/L6}{\JournalTitle{\apjl},
  739, L6}

\bibitem[{{Li} {et~al.}(2017)}]{Li+17}
{Li}, K.-L., {et~al.} 2017,
  \href{http://dx.doi.org/10.1038/s41550-017-0222-1}{\JournalTitle{Nature
  Astronomy}, 1, 697}

\bibitem[{{Linford} {et~al.}(2015){Linford}, {Ribeiro}, {Chomiuk}, {Nelson},
  {Sokoloski}, {Rupen}, {Mukai}, {O'Brien}, {Mioduszewski}, \&
  {Weston}}]{Linford+15}
{Linford}, J.~D., {Ribeiro}, V.~A.~R.~M., {Chomiuk}, L., {et~al.} 2015,
  \href{http://dx.doi.org/10.1088/0004-637X/805/2/136}{\JournalTitle{\apj},
  805, 136}

\bibitem[{{Livio} {et~al.}(1990){Livio}, {Shankar}, {Burkert}, \&
  {Truran}}]{Livio+90}
{Livio}, M., {Shankar}, A., {Burkert}, A., \& {Truran}, J.~W. 1990,
  \href{http://dx.doi.org/10.1086/168836}{\JournalTitle{\apj}, 356, 250}

\bibitem[{{Lloyd} {et~al.}(1997){Lloyd}, {O'Brien}, \& {Bode}}]{Lloyd+97}
{Lloyd}, H.~M., {O'Brien}, T.~J., \& {Bode}, M.~F. 1997,
  \href{http://dx.doi.org/10.1093/mnras/284.1.137}{\JournalTitle{\mnras}, 284,
  137}

\bibitem[{{Martin} \& {Dubus}(2013)}]{Martin&Dubus13}
{Martin}, P., \& {Dubus}, G. 2013,
  \href{http://dx.doi.org/10.1051/0004-6361/201220289}{\JournalTitle{\aap},
  551, A37}

\bibitem[{{Martin} {et~al.}(2018){Martin}, {Dubus}, {Jean}, {Tatischeff}, \&
  {Dosne}}]{Martin+18}
{Martin}, P., {Dubus}, G., {Jean}, P., {Tatischeff}, V., \& {Dosne}, C. 2018,
  \href{http://dx.doi.org/10.1051/0004-6361/201731692}{\JournalTitle{\aap},
  612, A38}

\bibitem[{{Mason} {et~al.}(2018){Mason}, {Shore}, {De Gennaro Aquino}, {Izzo},
  {Page}, \& {Schwarz}}]{Mason+18}
{Mason}, E., {Shore}, S.~N., {De Gennaro Aquino}, I., {et~al.} 2018,
  \href{http://dx.doi.org/10.3847/1538-4357/aaa247}{\JournalTitle{\apj}, 853,
  27}

\bibitem[{{Metzger} {et~al.}(2016){Metzger}, {Caprioli}, {Vurm}, {Beloborodov},
  {Bartos}, \& {Vlasov}}]{Metzger+16}
{Metzger}, B.~D., {Caprioli}, D., {Vurm}, I., {et~al.} 2016,
  \href{http://dx.doi.org/10.1093/mnras/stw123}{\JournalTitle{\mnras}, 457,
  1786}

\bibitem[{{Metzger} {et~al.}(2015){Metzger}, {Finzell}, {Vurm}, {Hasco{\"e}t},
  {Beloborodov}, \& {Chomiuk}}]{Metzger+15}
{Metzger}, B.~D., {Finzell}, T., {Vurm}, I., {et~al.} 2015,
  \href{http://dx.doi.org/10.1093/mnras/stv742}{\JournalTitle{\mnras}, 450,
  2739}

\bibitem[{{Metzger} {et~al.}(2014){Metzger}, {Hasco{\"e}t}, {Vurm},
  {Beloborodov}, {Chomiuk}, {Sokoloski}, \& {Nelson}}]{Metzger+14}
{Metzger}, B.~D., {Hasco{\"e}t}, R., {Vurm}, I., {et~al.} 2014,
  \href{http://dx.doi.org/10.1093/mnras/stu844}{\JournalTitle{\mnras}, 442,
  713}

\bibitem[{{Mioduszewski} \& {Rupen}(2004)}]{Mioduszewski&Rupen04}
{Mioduszewski}, A.~J., \& {Rupen}, M.~P. 2004,
  \href{http://dx.doi.org/10.1086/424376}{\JournalTitle{\apj}, 615, 432}

\bibitem[{{Mukai} \& {Ishida}(2001)}]{Mukai&Ishida01}
{Mukai}, K., \& {Ishida}, M. 2001,
  \href{http://dx.doi.org/10.1086/320220}{\JournalTitle{\apj}, 551, 1024}

\bibitem[{{Mukai} {et~al.}(2008){Mukai}, {Orio}, \& {Della Valle}}]{Mukai+08}
{Mukai}, K., {Orio}, M., \& {Della Valle}, M. 2008,
  \href{http://dx.doi.org/10.1063/1.2945023}{in American Institute of Physics
  Conference Series, Vol. 1010, A Population Explosion: The Nature \& Evolution
  of X-ray Binaries in Diverse Environments, ed. R.~M. {Bandyopadhyay},
  S.~{Wachter}, D.~{Gelino}, \& C.~R. {Gelino}}, 143

\bibitem[{{Nelson} {et~al.}(2019){Nelson}, {Mukai}, {Li}, {Vurm}, {Metzger},
  {Chomiuk}, {Sokoloski}, {Linford}, {Bohlsen}, \& {Luckas}}]{Nelson+19}
{Nelson}, T., {Mukai}, K., {Li}, K.-L., {et~al.} 2019,
  \href{http://dx.doi.org/10.3847/1538-4357/aafb6d}{\JournalTitle{\apj}, 872,
  86}

\bibitem[{{O'Brien} {et~al.}(1994){O'Brien}, {Lloyd}, \& {Bode}}]{OBrien+94}
{O'Brien}, T.~J., {Lloyd}, H.~M., \& {Bode}, M.~F. 1994,
  \href{http://dx.doi.org/10.1093/mnras/271.1.155}{\JournalTitle{\mnras}, 271,
  155}

\bibitem[{{Orio} {et~al.}(2001{\natexlab{a}}){Orio}, {Covington}, \&
  {{\"O}gelman}}]{Orio+01b}
{Orio}, M., {Covington}, J., \& {{\"O}gelman}, H. 2001{\natexlab{a}},
  \href{http://dx.doi.org/10.1051/0004-6361:20010537}{\JournalTitle{\aap}, 373,
  542}

\bibitem[{{Orio} {et~al.}(2001{\natexlab{b}}){Orio}, {Parmar}, {Benjamin},
  {Amati}, {Frontera}, {Greiner}, {{\"O}gelman}, {Mineo}, {Starrfield}, \&
  {Trussoni}}]{Orio+01a}
{Orio}, M., {Parmar}, A., {Benjamin}, R., {et~al.} 2001{\natexlab{b}},
  \href{http://dx.doi.org/10.1046/j.1365-8711.2001.04448.x}{\JournalTitle{\mnras},
  326, L13}

\bibitem[{{Osterbrock} \& {Ferland}(2006)}]{Osterbrock2006}
{Osterbrock}, D.~E., \& {Ferland}, G.~J. 2006, {Astrophysics of gaseous nebulae
  and active galactic nuclei}

\bibitem[{{Pejcha} {et~al.}(2016{\natexlab{a}}){Pejcha}, {Metzger}, \&
  {Tomida}}]{Pejcha+16b}
{Pejcha}, O., {Metzger}, B.~D., \& {Tomida}, K. 2016{\natexlab{a}},
  \href{http://dx.doi.org/10.1093/mnras/stw1481}{\JournalTitle{\mnras}, 461,
  2527}

\bibitem[{{Pejcha} {et~al.}(2016{\natexlab{b}}){Pejcha}, {Metzger}, \&
  {Tomida}}]{Pejcha+16a}
---. 2016{\natexlab{b}},
  \href{http://dx.doi.org/10.1093/mnras/stv2592}{\JournalTitle{\mnras}, 455,
  4351}

\bibitem[{{Pinto} \& {Eastman}(2000)}]{Pinto&Eastman00}
{Pinto}, P.~A., \& {Eastman}, R.~G. 2000,
  \href{http://dx.doi.org/10.1086/308380}{\JournalTitle{\apj}, 530, 757}

\bibitem[{{Pontefract} \& {Rawlings}(2004)}]{Pontefract&Rawlings04}
{Pontefract}, M., \& {Rawlings}, J.~M.~C. 2004,
  \href{http://dx.doi.org/10.1111/j.1365-2966.2004.07330.x}{\JournalTitle{\mnras},
  347, 1294}

\bibitem[{{Prialnik}(1986)}]{Prialnik86}
{Prialnik}, D. 1986,
  \href{http://dx.doi.org/10.1086/164677}{\JournalTitle{\apj}, 310, 222}

\bibitem[{{Quataert} {et~al.}(2016){Quataert}, {Fern{\'a}ndez}, {Kasen},
  {Klion}, \& {Paxton}}]{Quataert+16}
{Quataert}, E., {Fern{\'a}ndez}, R., {Kasen}, D., {Klion}, H., \& {Paxton}, B.
  2016, \href{http://dx.doi.org/10.1093/mnras/stw365}{\JournalTitle{\mnras},
  458, 1214}

\bibitem[{{Ribeiro} {et~al.}(2013){Ribeiro}, {Munari}, \&
  {Valisa}}]{Ribeiro+13}
{Ribeiro}, V.~A.~R.~M., {Munari}, U., \& {Valisa}, P. 2013,
  \href{http://dx.doi.org/10.1088/0004-637X/768/1/49}{\JournalTitle{\apj}, 768,
  49}

\bibitem[{{Ribeiro} {et~al.}(2014{\natexlab{a}}){Ribeiro}, {Chomiuk}, {Munari},
  {Steffen}, {Koning}, {O'Brien}, {Simon}, {Woudt}, \& {Bode}}]{Ribereiro+14}
{Ribeiro}, V.~A.~R.~M., {Chomiuk}, L., {Munari}, U., {et~al.}
  2014{\natexlab{a}},
  \href{http://dx.doi.org/10.1088/0004-637X/792/1/57}{\JournalTitle{\apj}, 792,
  57}

\bibitem[{{Ribeiro} {et~al.}(2014{\natexlab{b}}){Ribeiro}, {Chomiuk}, {Munari},
  {Steffen}, {Koning}, {O'Brien}, {Simon}, {Woudt}, \& {Bode}}]{Ribeiro+14}
---. 2014{\natexlab{b}},
  \href{http://dx.doi.org/10.1088/0004-637X/792/1/57}{\JournalTitle{\apj}, 792,
  57}

\bibitem[{{Roy} {et~al.}(2012){Roy}, {Chomiuk}, {Sokoloski}, {Weston}, {Rupen},
  {Johnson}, {Krauss}, {Nelson}, {Mukai}, \& {Mioduszewski}}]{Roy+12}
{Roy}, N., {Chomiuk}, L., {Sokoloski}, J.~L., {et~al.} 2012,
  \JournalTitle{Bulletin of the Astronomical Society of India}, 40, 293

\bibitem[{{Rybicki} \& {Lightman}(1979)}]{Rybicki&Lightman79}
{Rybicki}, G.~B., \& {Lightman}, A.~P. 1979, {Radiative processes in
  astrophysics}

\bibitem[{{Schwarz} {et~al.}(2011){Schwarz}, {Ness}, {Osborne}, {Page},
  {Evans}, {Beardmore}, {Walter}, {Helton}, {Woodward}, {Bode}, {Starrfield},
  \& {Drake}}]{Schwarz+11}
{Schwarz}, G.~J., {Ness}, J.-U., {Osborne}, J.~P., {et~al.} 2011,
  \href{http://dx.doi.org/10.1088/0067-0049/197/2/31}{\JournalTitle{\apjs},
  197, 31}

\bibitem[{{Shankar} {et~al.}(1991){Shankar}, {Livio}, \& {Truran}}]{Shankar+91}
{Shankar}, A., {Livio}, M., \& {Truran}, J.~W. 1991,
  \href{http://dx.doi.org/10.1086/170148}{\JournalTitle{\apj}, 374, 623}

\bibitem[{{Shore}(2013)}]{Shore13}
{Shore}, S.~N. 2013,
  \href{http://dx.doi.org/10.1051/0004-6361/201322470}{\JournalTitle{\aap},
  559, L7}

\bibitem[{{Shore} {et~al.}(2013){Shore}, {De Gennaro Aquino}, {Schwarz},
  {Augusteijn}, {Cheung}, {Walter}, \& {Starrfield}}]{Shore+13}
{Shore}, S.~N., {De Gennaro Aquino}, I., {Schwarz}, G.~J., {et~al.} 2013,
  \href{http://dx.doi.org/10.1051/0004-6361/201321095}{\JournalTitle{\aap},
  553, A123}

\bibitem[{{Sokoloski} {et~al.}(2008){Sokoloski}, {Rupen}, \&
  {Mioduszewski}}]{Sokoloski+08}
{Sokoloski}, J.~L., {Rupen}, M.~P., \& {Mioduszewski}, A.~J. 2008,
  \href{http://dx.doi.org/10.1086/592602}{\JournalTitle{\apj}, 685, L137}

\bibitem[{{Starrfield} {et~al.}(2016){Starrfield}, {Iliadis}, \&
  {Hix}}]{Starrfield+16}
{Starrfield}, S., {Iliadis}, C., \& {Hix}, W.~R. 2016,
  \href{http://dx.doi.org/10.1088/1538-3873/128/963/051001}{\JournalTitle{\pasp},
  128, 051001}

\bibitem[{{Starrfield} {et~al.}(1972){Starrfield}, {Truran}, {Sparks}, \&
  {Kutter}}]{Starrfield+72}
{Starrfield}, S., {Truran}, J.~W., {Sparks}, W.~M., \& {Kutter}, G.~S. 1972,
  \href{http://dx.doi.org/10.1086/151619}{\JournalTitle{\apj}, 176, 169}

\bibitem[{{Steinberg} \& {Metzger}(2018)}]{Steinberg&Metzger18}
{Steinberg}, E., \& {Metzger}, B.~D. 2018,
  \href{http://dx.doi.org/10.1093/mnras/sty1641}{\JournalTitle{\mnras}, 479,
  687}

\bibitem[{{Strope} {et~al.}(2010){Strope}, {Schaefer}, \& {Henden}}]{Strope+10}
{Strope}, R.~J., {Schaefer}, B.~E., \& {Henden}, A.~A. 2010,
  \href{http://dx.doi.org/10.1088/0004-6256/140/1/34}{\JournalTitle{\aj}, 140,
  34}

\bibitem[{{Tanaka} {et~al.}(2011{\natexlab{a}}){Tanaka}, {Nogami}, {Fujii},
  {Ayani}, \& {Kato}}]{Tanaka+11a}
{Tanaka}, J., {Nogami}, D., {Fujii}, M., {Ayani}, K., \& {Kato}, T.
  2011{\natexlab{a}},
  \href{http://dx.doi.org/10.1093/pasj/63.1.159}{\JournalTitle{\pasj}, 63, 159}

\bibitem[{{Tanaka} {et~al.}(2011{\natexlab{b}}){Tanaka}, {Nogami}, {Fujii},
  {Ayani}, {Kato}, {Maehara}, {Kiyota}, \& {Nakajima}}]{Tanaka+11b}
{Tanaka}, J., {Nogami}, D., {Fujii}, M., {et~al.} 2011{\natexlab{b}},
  \href{http://dx.doi.org/10.1093/pasj/63.4.911}{\JournalTitle{\pasj}, 63, 911}

\bibitem[{{Townsend}(2009)}]{Townsend}
{Townsend}, R.~H.~D. 2009,
  \href{http://dx.doi.org/10.1088/0067-0049/181/2/391}{\JournalTitle{\apjs},
  181, 391}

\bibitem[{{Vishniac}(1994)}]{Vishniac94}
{Vishniac}, E.~T. 1994,
  \href{http://dx.doi.org/10.1086/174231}{\JournalTitle{\apj}, 428, 186}

\bibitem[{{Vlasov} {et~al.}(2016){Vlasov}, {Vurm}, \& {Metzger}}]{Vlasov+16}
{Vlasov}, A., {Vurm}, I., \& {Metzger}, B.~D. 2016,
  \href{http://dx.doi.org/10.1093/mnras/stw1949}{\JournalTitle{\mnras}, 463,
  394}

\bibitem[{{Vurm} \& {Metzger}(2018)}]{Vurm&Metzger18}
{Vurm}, I., \& {Metzger}, B.~D. 2018,
  \href{http://dx.doi.org/10.3847/1538-4357/aa9c4a}{\JournalTitle{\apj}, 852,
  62}

\bibitem[{{Walter} {et~al.}(2012){Walter}, {Battisti}, {Towers}, {Bond}, \&
  {Stringfellow}}]{Walter+12}
{Walter}, F.~M., {Battisti}, A., {Towers}, S.~E., {Bond}, H.~E., \&
  {Stringfellow}, G.~S. 2012,
  \href{http://dx.doi.org/10.1086/668404}{\JournalTitle{\pasp}, 124, 1057}

\bibitem[{{Weston} {et~al.}(2016){Weston}, {Sokoloski}, {Chomiuk}, {Linford},
  {Nelson}, {Mukai}, {Finzell}, {Mioduszewski}, {Rupen}, \&
  {Walter}}]{Weston+16}
{Weston}, J.~H.~S., {Sokoloski}, J.~L., {Chomiuk}, L., {et~al.} 2016,
  \href{http://dx.doi.org/10.1093/mnras/stw1161}{\JournalTitle{\mnras}, 460,
  2687}

\bibitem[{{Williams} {et~al.}(2008){Williams}, {Mason}, {Della Valle}, \&
  {Ederoclite}}]{Williams+08}
{Williams}, R., {Mason}, E., {Della Valle}, M., \& {Ederoclite}, A. 2008,
  \href{http://dx.doi.org/10.1086/590056}{\JournalTitle{\apj}, 685, 451}

\bibitem[{{Wolf} {et~al.}(2013){Wolf}, {Bildsten}, {Brooks}, \&
  {Paxton}}]{Wolf+13}
{Wolf}, W.~M., {Bildsten}, L., {Brooks}, J., \& {Paxton}, B. 2013,
  \href{http://dx.doi.org/10.1088/0004-637X/777/2/136}{\JournalTitle{\apj},
  777, 136}

\bibitem[{{Woudt} {et~al.}(2009){Woudt}, {Steeghs}, {Karovska}, {Warner},
  {Groot}, {Nelemans}, {Roelofs}, {Marsh}, {Nagayama}, \& {Smits}}]{Woudt+09}
{Woudt}, P.~A., {Steeghs}, D., {Karovska}, M., {et~al.} 2009,
  \href{http://dx.doi.org/10.1088/0004-637X/706/1/738}{\JournalTitle{\apj},
  706, 738}

\bibitem[{{Yalinewich} {et~al.}(2015){Yalinewich}, {Steinberg}, \&
  {Sari}}]{Yalinewich+15}
{Yalinewich}, A., {Steinberg}, E., \& {Sari}, R. 2015,
  \href{http://dx.doi.org/10.1088/0067-0049/216/2/35}{\JournalTitle{\apjs},
  216, 35}

\bibitem[{{Yaron} {et~al.}(2005){Yaron}, {Prialnik}, {Shara}, \&
  {Kovetz}}]{Yaron+05}
{Yaron}, O., {Prialnik}, D., {Shara}, M.~M., \& {Kovetz}, A. 2005,
  \href{http://dx.doi.org/10.1086/428435}{\JournalTitle{\apj}, 623, 398}

\bibitem[{{Zdziarski} \& {Svensson}(1989)}]{Zdziarski&Svensson89}
{Zdziarski}, A.~A., \& {Svensson}, R. 1989,
  \href{http://dx.doi.org/10.1086/167826}{\JournalTitle{\apj}, 344, 551}

\end{thebibliography}

\bibliographystyle{yahapj}







\appendix

\appendix
\section{Numerical Method}
\label{app:numerical}
We solve the Euler equations using a 1D second-order Lagrangian Finite-Volume scheme with spherical geometry. The code employs an exact Riemann solver with an ideal gas equation of state with a polytropic index $\gamma=5/3$. The code has been validated against a suite of smooth (e.g. linear waves) and non-smooth (e.g. Sedov blast wave) test problems. 

During the run we attempt to enforce cells to have a size $dr\approx 3\cdot 10^{-4}r$. However, once shocked cells cool, their size decreases by a factor of $\sim \mathcal{M}^2$, which greatly reduces the time step. Whenever a cell is smaller than the desired threshold, we merge it with its smaller neighbor, provided that it is in a smooth region. If the cell is not in a smooth region, we remove it only if it satisfies that the ratio in the pressure, temperature and density between both of its neighbors satisfies:
\begin{eqnarray}
p_\textrm{ratio}&<&1.1\cdot(1.5\cdot10^{-4}r/dr)^{0.9}\\
T_\textrm{ratio}&<&1.2\cdot(1.5\cdot10^{-4}r/dr)^{0.1}\\
\rho_\textrm{ratio}&<&1.2\cdot(1.5\cdot10^{-4}r/dr)^{1.1}
\end{eqnarray}

We include optically-thin radiative losses using the cooling function calculated from version C17.01 of CLOUDY \citep{Ferland+98,Ferland+17} using the GASS10 (solar) abundances and the exact cooling integration method as described in \citet{Townsend}.
\section{Radio Opacity}
\label{app:radio}
In this section we outline our algorithm to determine the radio opacity.
We start by identifying the location of the outermost shock, $r_{\rm f}$. 
We assume that the WD has an ionizing photon luminosity of 
\begin{equation}
    Q_\textrm{h,WD}=L_\textrm{WD}/13.6\,\textrm{eV}\;{\rm s^{-1}},
\end{equation}
and we subtract from it the total rate of recombinations up to the forward shock
\begin{equation}
     Q_{\rm h,net} = {\rm max}\left(Q_\textrm{h,WD}-\int_0^{r_{\rm f}}\alpha_{\rm B}{\rm n_e n_i 4\pi r^2 dr} ,0\right),
\end{equation}
where $\alpha_{\rm B} = 2.6\cdot10^{-13}\left(\frac{T}{10^4\,{\rm K}}\right)^{-0.8}\,{\rm cm^3\,s^{-1}}$ is the case B recombination rate \citep{Osterbrock2006}. The final ionizing photon flux emerging from the forward shock, $Q_{\rm h}$ is the net transmitted ionizing photon flux, $ Q_{\rm h,net} $, with the addition of the ionizing photon flux generated at the forward shock assuming a free-free spectrum. Following the procedure described in \cite{Metzger+14} (eq. 42 and 43) we estimate the ionization state of the hydrogen above the forward shock.

If the radio photosphere is above the forward shock (i.e. $r(\tau=1)>r_{\rm f}$), we estimate the free-free radio emission to be
\begin{equation}
    \nu L_{\nu} = 4\pi^{2}R_{\rm ph,\nu}^{2}\frac{2\nu^{3} kT^{\rm ion}}{c^{2}} +
    \int^\infty_{r(\tau=1)} \varepsilon_{ff} 4\pi r^2 dr,
\end{equation}
where $ \varepsilon_{ff}$ is the free-free emissivity and $T^{\rm ion}=10^4$K.

If the photosphere is below the forward shock, we calculate the ionization state of the gas between the surface of the Str{\" o}mgren sphere and the photosphere using the values for CIE based taken from \cite{Gnat07}.


\bsp	
\label{lastpage}
\end{document}